\documentstyle[aps,twocolumn]{revtex}
\begin{document}

\title{Statistical Theory of Finite Fermi-Systems\\
Based on the Structure of Chaotic Eigenstates}

\author{V.V.Flambaum$^{1,2}$ and 
F.M.Izrailev$^{1,2}$\thanks{email address: izrailev@physics.spa.umn.edu}}

\address{$^1$ School of Physics, University of New South Wales,
Sydney 2052, Australia}

\address{$^2$ Budker Institute of Nuclear Physics, 630090 Novosibirsk, Russia}

\twocolumn[
\date{\today}

\maketitle

\widetext

\vspace*{-1.0truecm}

\begin{abstract}
\parbox{14cm}

The approach is developed for the description of isolated Fermi-systems with
finite number of particles, such as complex atoms, nuclei, atomic clusters
etc. It is based on statistical properties of chaotic excited states which
are formed by the interaction between particles. New type of
``microcanonical'' partition function is introduced and expressed in terms of
the average shape of eigenstates $F(E_k,E)$ where $E$ is the
total energy of the system. This partition function plays the
same role as the canonical expression $exp(-E^{(i)}/T)$ for open
systems in thermal bath.
The approach allows to calculate mean values
and non-diagonal matrix elements of different operators. In particular, the
following problems have been considered: distribution of occupation numbers
and its relevance to the canonical and Fermi-Dirac distributions; criteria
of equilibrium and thermalization; thermodynamical equation of state and the
meaning of temperature, entropy and heat capacity, increase of effective
temperature due to the interaction. The problems of spreading
widths and shape of the eigenstates are also studied. 
\end{abstract}

\pacs{PACS numbers:  05.45.+b, 31.25.-v, 31.50.+w, 32.30.-r}

] \narrowtext

\section{Introduction}

As is known, quantum statistical laws are derived for systems with infinite
number of particles, or for systems in a heat bath, therefore, their
applicability to isolated finite systems of a few particles is, at least,
questionable. However, the density of many-particle energy levels increases
extremely fast (typically, exponentially) both with an increase of number of
particles and excitation energy. For this reason, even a weak interaction
between particles can lead to a strong mixing between large number of simple
many-particle states, resulting in the so-called {\em chaotic eigenstates. }%
If the components of such eigenstates can be treated as random variables
(onset of {\em Quantum Chaos} ), $\,$statistical methods are expected to be
valid even for an isolated dynamical system.

One should stress that statistical description of such isolated systems can
be quite different from that based on standard canonical distributions;
therefore, application of the famous Fermi-Dirac or Bose-Einstein formulae
may give incorrect results. Moreover, for isolated few-particle systems a
serious problem arises in the definition of temperature, or other
thermodynamic variables like entropy and specific heat (to compare with, in
infinite systems different definitions give the same result).

The aim of this paper is to develop statistical theory for finite quantum
systems of interacting particles, based on generic statistical properties of
chaotic eigenstates (``microcanonical'' approach). Typical examples of such
systems are compound nuclei, complex atoms, atomic clusters, isolated
quantum dots, etc.

The structure of the paper is as follows. In the next Section 2, new type of
``microcanonical'' partition function is introduced for finite isolated
systems, which is directly related to the average shape of chaotic
eigenstates. Based on this partition function, general expression for the
occupation numbers is given which is valid for any number of interacting
particles. Relation of this ``microcanonical'' expression ($F-{\em %
distribution}$) to that of standard canonical distribution for occupation
numbers is the subject of the Section 3. Also, here a new form of the
canonical distribution is suggested which can be convenient in applications
to open systems in thermal equilibrium like quantum dots.

Transition to the standard Fermi-Dirac distribution for large systems is
analyzed in Section 4. Specific attention is paid to the accuracy of this
distribution in applications to isolated systems with few particles.

In Section 5 the analytical derivation of the 
$F-$distribution for occupation
numbers is given. For this, the model of $n$ randomly interacting
Fermi-particles distributed over $m$ single-particle levels has been used.
The analytical formula for the occupation numbers $n_s(E)$ with $E$ being
the total energy of the system, was found to be in excellent agreement with
the numerical experiment.

In Section 6 the influence of finite width of single-particles states
(``quasi-particles'') on the distribution of occupation numbers is
considered. New expression for the occupation numbers is discussed which
takes into account the finite spreading width of ``quasi-particles''. It is
demonstrated that for an isolated system with the fixed total energy $E$ ,
the incorporation of spreading widths decreases the effective temperature,
however, does not change the occupation numbers.

Section 7 deals with an important problem of thermodynamical description of
small systems consisting of finite number of interacting particles.
Specifically, different definitions of both temperature and entropy have
been analyzed, and the equation of state for finite systems has been derived.

In Section 8 we show that statistical effects of the interaction can be
imitated by an increase of the effective temperature. This fact allows to
use the standard Fermi-Dirac expression for the occupation numbers (with
renormalized parameters) in the application to both isolated and open (in
the thermal bath) systems of interacting Fermi-particles. This interaction
may be strong, however, the assumption of randomness for residual
interaction matrix elements is essential.

In Section 9 the conditions for chaos, equilibrium and ``thermalization''
have been analyzed for an isolated quantum system of a finite number of
particles in terms of the interaction strength, number of particles and
other related parameters. Depending on these conditions, there are four
different regions with different distributions of occupation numbers, which
are discussed in detail. In this Section, the transition to mesoscopic
systems is shortly discussed. The question of particular interest is how
statistical properties depend on the dimensionality of a system.

Since the approach developed in this paper is entirely related to the
structure of chaotic eigenstates, in Appendix 1 the analytical expression
for the average shape of eigenstates is given and discussed in more details.
This expression, which also describes the local spectral density of states
(LDOS), essentially depends on two different widths and is valid both for
weak and strong interaction between particles. For weak interaction the
shape is close to the Lorentzian form with the half-width given by the Fermi
Golden Rule. For larger interaction the shape is characterized by
exponential tails and by the width which is linear in the interaction
strength (contrary to the Fermi Golden Rule which gives a quadratic
dependence). Transition between these two regimes occurs when the half-width
is comparable to the root-mean-square width (effective band width of the
Hamiltonian matrix).

In Appendix 2 the moments of the distributions of the basis components
(LDOS) and energy levels are calculated.

And finally, in Appendix 3 the calculation of density of final states and
spreading widths of the LDOS has been performed using the Fermi Golden Rule.

\section{Microcanonical partition function}

In this Section we are going to derive the partition function for closed
(isolated) systems of finite number of interacting particles. This function
allows one to perform analytical and numerical calculations of statistical
mean values of different operators, for example, occupation numbers.

We follow the standard approach which is based on the separation of a total
Hamiltonian in two parts, 
\begin{equation}
\label{H}H=H_0+V=\sum \epsilon _sa_s^{+}a_s+\frac 12\sum
V_{pqrs}a_p^{+}a_q^{+}a_ra_s 
\end{equation}
The ``unperturbed'' Hamiltonian $H_0$ should incorporate the effect of a
mean field (if any), $\epsilon _s$ are the energies of single-particle
states (``orbitals'') calculated in this field, $a_s^{+},a_s$ are
creation-annihilation operators, and $V$ stands for the residual
interaction. For simplicity, here we neglect any dynamical effects of the
interaction like pairing, collective modes etc. Instead, we will study
statistical effects of interaction, therefore, in what follows we assume
that matrix elements $V_{pqrs}$ of the two-body residual interaction are
random variables.

Exact (``compound'') eigenstates $\left| i\right\rangle \,$of the
Hamiltonian $H$ can be expressed in terms of simple ``shell-model basis
states'' $\left| k\right\rangle \,$(eigenstates of $H_0$) :

\begin{equation}
\label{slat}\left| i\right\rangle =\sum\limits_kC_k^{(i)}\left|
k\right\rangle \,;\,\,\,\,\,\,\,\,\left| k\right\rangle
=a_{k_1}^{+}...a_{k_n}^{+}\left| 0\right\rangle 
\end{equation}
These compound eigenstates $\left| i\right\rangle $ , characterized by the
corresponding energies $E^{(i)}$ , are formed by the residual interaction $V$
; in complex systems they typically contain large number $N_{pc}\gg 1$ of
the so-called principal components $C_k^{(i)}$ which fluctuate ``randomly''
as a function of indices $i$ and $k$ .

Our main interest is in the occupation numbers $n_s$ of single-particle
states (orbitals). They can be represented in terms of components of the
exact eigenstates $\left| i\right\rangle $ , 
\begin{equation}
\label{ns}n_s=\left\langle i\right| \hat n_s\left| i\right\rangle
=\sum\limits_k\left| C_k^{(i)}\right| ^2\left\langle k\right| \hat n_s\left|
k\right\rangle 
\end{equation}
Here $\hat n_s=a_s^{+}a_s$ is the occupation number operator. The knowledge
of the distribution of occupation numbers gives the possibility to calculate
mean values of any single particle operator $\left\langle M\right\rangle
=\sum_sn_sM_{ss}$ . Moreover, the variance of the distribution of
non-diagonal elements of $M$ describing transition amplitudes between
``chaotic'' compound states, can be also expressed through the occupation
numbers $n_s$ \cite{F,FGGK94,FGI96} .

As one can see from (\ref{ns}), mean values of occupation numbers depend on
the shape of exact eigenstates, given by the ``spreading function'' $F$ (in
what follows, the $F-{\em function}$ ), 
\begin{equation}
\label{Fk}F_k^{(i)}\equiv \overline{\left| C_k^{(i)}\right| ^2}%
=F\,(E_k-E^{(i)}) 
\end{equation}
Last equality in the above expression reflects the fact that residual
interaction $V$ mainly mixes close components $k$ in some energy interval $%
\Gamma $ (``spreading width'') nearby the eigenstate energy $E^{(i)}$ (more
accurately, nearby the unperturbed energy $E_k$ for $k=i$ , see below) .

Typically, this spreading function rapidly decreases with an increase of $%
\left| E_k-E^{(i)}\right| $ (since an admixture of distant component is very
small). Recently, in numerical studies of the Ce atom \cite{FGGK94}, the $%
s-d $ nuclear shell model \cite{ZELE} and random two-body interaction model 
\cite{FGI96,FIC96} it was found that typical shape $F$ of exact eigenstates
practically does not depend on a particular many-body system and has a
universal form which essentially depends on the spreading width $\Gamma $ .
The latter can be expressed in terms of parameters of the model ( intensity $%
V$ of the residual interaction, number of particles $n$ , excitation energy,
etc.) and can be calculated analytically (see Appendices 1- 3). One can also
measure the width of the $F-$function (\ref{Fk}) via the number of principal
components, $N_{pc}\sim \Gamma /D$ where $D\,$ is the local mean energy
spacing for compound states. In many-body systems the value of $D\,$
exponentially decreases with an increase of the number of ``active''
(valence) particles. As a result, $N_{pc}\,$ is very large, $\sim 10^4\div
10^6$ in excited (compound) nuclei and $\sim 100\,$ in excited rare-earth or
actinide atoms.

The starting point of our approach is the expression for the occupation
numbers which stems from Eqs. ($\ref{ns},\ref{Fk}),$ 
\begin{equation}
\label{nalpha}n_s(E)=\frac{\sum\limits_kn_s^{(k)}\,F(E_k-E)}{%
\sum\limits_k\,F\,(E_k-E)} 
\end{equation}
where $n_s^{(k)}\equiv \left\langle k\left| \hat n_s\right| k\right\rangle $
equal $0$ or $1$ for Fermi-particles and the sum in the denominator stands
for the normalization$.$ This way of averaging of occupation numbers is a
kind of microcanonical averaging since it is defined for the fixed total
energy $E$ of a system. One can see that, in fact, the relation (\ref{nalpha}%
) is equivalent to the introduction of a new kind of partition function, 
\begin{equation}
\label{ZZ}Z(E)=\sum_kF(E_k-E) 
\end{equation}
which is entirely determined by the shape of chaotic eigenfunctions. In what
follows, we term Eq.(\ref{nalpha}) the $F-{\em distribution}$.

The above expression (\ref{nalpha}) gives a new insight on the problem of
statistical description of complex systems. Indeed, as was mentioned above,
the shape of the $F-$ function has universal features and can be often
described analytically, therefore, in practice there is no need to
diagonalize huge Hamiltonian matrix of many-body system in order to find
statistical averages. One should stress that the summation in (\ref{nalpha})
is carried out over unperturbed energies $E_k$ defined by the mean field,
rather than over the energies of exact eigenstates in the standard canonical
distribution. As a result, the distribution of occupation numbers can be
derived analytically (see Section 5) even for few interacting particles, in
the situation when the standard Fermi-Dirac distribution is not valid.

\section{Transition to the canonical distribution}

It is instructive to compare our $F-$ distribution (\ref{nalpha}) with
occupation numbers obtained by making use of the standard canonical
distribution, 
\begin{equation}
\label{gibbs}n_s(T)=\frac{\sum\limits_in_s^{(i)}\,\exp (-E^{(i)}/T)}{%
\sum\limits_i\exp (-E^{(i)}/T)} 
\end{equation}
where $T$ is the temperature and the index $i$ stands for exact eigenstates.
The important difference between the $F-$ distribution (\ref{nalpha}) and
the canonical distribution (\ref{gibbs}) is that in Eq. (\ref{nalpha}) the
occupation numbers are calculated for a specific energy $E$ of a system
unlike specific temperature $T$ in Eq.(\ref{gibbs}). However, results of
calculations based on Eqs. (\ref{gibbs}) and (\ref{nalpha}) can be compared
with each other using the relation between the energy $E$ and the
temperature $T$ , 
\begin{equation}
\label{energy}E=\left\langle E\right\rangle _T=\frac{\sum\limits_iE^{(i)}\,%
\exp (-E^{(i)}/T)}{\sum\limits_i\exp (-E^{(i)}/T)} 
\end{equation}
The comparison of Eqs. (\ref{gibbs}) and (\ref{nalpha}) also shows that the
canonical distribution corresponds to the averaging of the
``microcanonical'' $F-$ distribution over some energy interval $\Delta _T$ .
To demonstrate this, let us substitute $n_s^{(i)}$ and $\overline{\left|
C_k^{(i)}\right| ^2}$ from Eqs.(\ref{ns},\ref{Fk}) into Eq.(\ref{gibbs}) and
replace the summation over $i$ by the integration over $\rho
(E^{(i)})dE^{(i)}$ where $\rho (E^{(i)})$ is the density of exact energy
levels, 
\begin{equation}
\label{Ffi}\sum_in_s^{(i)}\exp \left( -E^{(i)}/T\right) \approx \int
n_s^{(i)}\Phi _T(E^{(i)})dE^{(i)} 
\end{equation}
Here we have introduced the ``canonical (thermal) averaging'' function, 
\begin{equation}
\label{fi}\Phi _T(E)=\rho (E)\exp \left( -E/T\right) 
\end{equation}
which is discussed below. As a result, we can transform canonical
distribution (\ref{gibbs}) into the form similar to the $F-$ distribution (%
\ref{nalpha}), 
\begin{equation}
\label{nsT}n_s(T)=\frac{\sum\limits_kn_s^{(k)}\,F(T,E_k)}{%
\sum\limits_k\,F\,(T,E_k)} 
\end{equation}
where the function $F(T,E_k)$ is the canonical average of $F_k^{(i)}$ , 
\begin{equation}
\label{FT}F(T,E_k)=\int F_k^{(i)}\Phi _T(E^{(i)})\,dE^{(i)} 
\end{equation}

Note that this form of the canonical distribution can be convenient for the
calculation of the occupation numbers and other mean values in quantum dots
which are in thermal equilibrium with an environment (with no particle
exchange).

In large many-body systems the canonical averaging function $\Phi _T(E)$ has
narrow maximum since the density of states $\rho (E^{(i)})\,$ typically
grows very fast. The position $E_m$ of its maximum is defined by the
expression 
\begin{equation}
\label{Tt}\frac{d\ln \rho (E)}{dE}=\frac 1T 
\end{equation}
and the width is given by 
\begin{equation}
\label{deltaT}\Delta _T=\left| \frac{d^2\ln \rho (E)}{dE^2}\right| ^{-1/2} 
\end{equation}
As an example, let us consider the system of $n$ interacting particles
distributed over $m$ orbitals. In the papers \cite{FW70,BF71} it was shown
that in the case $m\gg n\gg 1$ the density of states is of the Gaussian
form, 
\begin{equation}
\label{rho}\rho (E)=\frac 1{\sigma \sqrt{2\pi }}\exp \left( -\frac{\left(
E-E_c\right) ^2}{2\sigma ^2}\right) 
\end{equation}
where $E_c$ is the center of the spectrum and $\sigma ^2$ is its variance.
According to recent numerical data \cite{FIC96,FGI96}, the Gaussian form for
the density $\rho (E)$ occurs also for few particles ($n\geq 4$) . This fact
allows easily to find the form of $\Phi _T(E)$ which appears to be of quite
generic, 
\begin{equation}
\label{Fform}\Phi _T(E) \sim \exp \left( -\frac{%
\left( E-E_m\right) ^2}{2\sigma ^2}\right) 
\end{equation}
where 
\begin{equation}
\label{Em}E_m=E_c-\frac{\sigma ^2}T 
\end{equation}
One can see that the width $\Delta _T$ of the thermal averaging function
equals to the root-mean--square (r.m.s.) width of the spectrum, $\Delta
_T=\sigma $ . Now, it is easy to show that the thermal averaging width $%
\Delta _T$ is always bigger than the r.m.s. width $\Delta E$ of the
``microcanonical'' $F-$ function or, the same, than the mean width of exact
eigenstates in the energy representation. Indeed, there is a simple relation
between the widths $\sigma $ and $\sigma _0$ of the energy spectrum with and
without interaction, respectively, (see Appendix 1), 
\begin{equation}
\label{sigma}\sigma ^2=\sigma _0^2+(\Delta E)^2\,\,, 
\end{equation}
therefore, we have $\Delta _T=\sigma \,>$ $\Delta E$ . One should stress
that the latter width $\Delta E$ , in fact, is due to statistical effects of
interaction. The difference between the widths $\Delta _T$ and $\Delta E$ is
not important when the number of particles $n$ is large. This is because
with an increase of $n$ the width $\sigma _0$ of the unperturbed spectrum
increases as $\sqrt{n}$, unlike the width $\Delta E$ which increases as $n$.
One should also note that in this case both widths $\Delta _T$ and $\Delta E$
are much smaller than the typical energy interval, $\sigma /\left|
E-E_c\right| \sim 1/\sqrt{n}$ . Therefore, for large number of particles the
function $\Phi _T$ can be regarded as the delta-function at $E=E_m$ and the $%
F-$ distribution is close to the canonical distribution, see Eq.(\ref{FT}).

To conclude with this Section, the canonical distribution (\ref{gibbs}) is
not correct when describing isolated systems with small number of particles,
instead, one should use the $F-$ distribution (\ref{nalpha}). This was
recently confirmed by numerical experiments with the model of few
Fermi-particles with two-body random interaction \cite{FIC96,FI97}.

\section{Transition to the Fermi-Dirac distribution}

It is instructive now to show how the standard Fermi-Dirac distribution
stems directly from the $F-$ distribution (\ref{nalpha}) in the limit of
large number of particles. By performing explicitly the summation over $%
n_s=0 $ and $n_s=1$ , the expression (\ref{nalpha}) can be rewritten in the
form 
\begin{equation}
\label{nsZ}n_s(E)=\frac{0+Z_s(n-1,E-\tilde \epsilon _s)}{Z_s(n-1,E-\tilde
\epsilon _s)+Z_s(n,E)}=\frac 1{1+\frac{Z_s(n,E)}{Z_s(n-1,E-\tilde \epsilon
_s)}} 
\end{equation}
Here, two ``partial'' partition functions $Z_s(n,E)$ and $Z_s(n-1,E-\tilde
\epsilon _s)$ are introduced. In the first one, the summation is taken over
all single-particle states of $n$ particles with the orbital $s$ excluded, $%
Z_s(n,E)=\sum\nolimits_k^{\prime }F(E_k-E)$. Correspondingly, the sum in $%
Z_s(n-1,E-\tilde \epsilon _s)$ is taken over the states of $n-1$ particles
with the orbital $s$ excluded. The latter sum appears from the terms for
which the orbital $s$ is filled $(n_s=1)$ , thus, we should add the energy $%
\tilde \epsilon _s\equiv E_k(n)-E_k(n-1)$ of this orbital to the energy $%
E_k(n-1)$ of the basis state $\left| k\right\rangle $ defined by $n-1$
particles. Since the $F-$ function depends on the difference $E_k+\tilde
\epsilon _s-E$ only, the adding term $\tilde \epsilon _s$ to $E_k(n-1)$ is
the same as its subtraction from the total energy $E$ . Note, that this term
is defined by

\begin{equation}
\label{eps}\tilde \epsilon _s=\epsilon _s+\sum\limits_{p\neq
s}u_{sp}n_p^{(k)} 
\end{equation}
where $\epsilon _s$ is the energy of a single-particle state and $u_{sp}$ is
the diagonal matrix element of the two-body residual interaction. By taking $%
\tilde \epsilon _s$ independent of $k$ we assume that the averaging over the
basis states near the energy $E$ is possible, in fact, this is equivalent to
a local (at a given energy) mean field approximation.

One should stress that this approximation is the most important when
applying the model (\ref{H}) to realistic systems. For example, for Ce atom
there are orbitals from different open sub-shells (e.g. $4f$ and $6s$ )
which are quite close in energies, however, they have very different radius.
As a result, the Coulomb interaction between the corresponding electrons is
very different \cite{FGGP97}. In this case the interaction terms in Eq.(\ref
{eps}) strongly depend on the occupation numbers of other particles, which
means that there is no good mean field approximation. As a result, the
equilibrium distribution for occupation numbers is very different from the
Fermi-Dirac distribution \cite{FGGP97}. However, the $F-$ distribution (\ref
{nalpha}) for occupation numbers is valid. In other cases like random
two-body interaction model \cite{FIC96,FGI96,FI97} or nuclear shell model 
\cite{ZELE}, such a local mean field approximation is quite accurate.

For large number $n\gg 1$ of particles distributed over $m\gg 1$ orbitals,
the dependence of $Z_s\,$ on $n$ and $\tilde \epsilon _s$ is very strong
since the number of terms $N$ in the partition function $Z_s\,$ is
exponentially large, $N=\frac{m!}{(m-n)!n!}$ . Therefore, to make the
dependence on arguments smooth, one should consider $\ln \,Z_s$ instead of $%
Z_s$ . In this case one can obtain 
$$
\ln \,Z_s(n-\Delta n,E-\tilde \epsilon _s)\approx \ln \,Z_s(n,E)-\alpha
_s\,\Delta n\,-\beta _s\tilde \epsilon _s 
$$
\begin{equation}
\label{ln}\alpha _s=\frac{\partial \ln \,Z_s}{\partial n};\,\,\,\,\,\,\beta
_s=\frac{\partial \ln \,Z_s}{\partial E};\,\,\,\,\,\,\Delta n=1
\end{equation}
This leads to the distribution of the Fermi-Dirac type, 
\begin{equation}
\label{FDtype}n_s=\frac 1{1+\exp (\alpha _s+\beta _s\tilde \epsilon _s)}
\end{equation}
If the number of substantially occupied orbitals in the definition of $Z_s\,$
is large, the parameters $\alpha _s$ and $\beta _s$ are not sensitive as to
which particular orbital $s$ is excluded from the sum and one can assume $%
\alpha _s=\alpha \equiv -\mu /T,\,\,\beta _s=\beta \equiv 1/T$ as in the
standard Fermi-Dirac distribution. Then, the chemical potential $\mu $ and
temperature $T$ can be found from the conditions of fixed number of
particles and fixed energy, 
\begin{equation}
\label{eqs}
\begin{array}{c}
\sum\limits_sn_s=n\, \\ 
\,\,\,\,\sum\limits_s\epsilon
_sn_s+\sum\limits_{s>p}u_{sp}n_sn_p=\sum\limits_sn_s(\epsilon _s+\tilde
\epsilon _s)/2=E
\end{array}
\end{equation}
Note, that the sums in (\ref{eqs} , \ref{eps}) containing the residual
interaction $u_{sp}$ can be substantially reduced by a proper choice of the
mean field basis (for instance, the terms $u_{sp}$ can have different signs
in such a basis). In practice, the values $\epsilon _s$ and $\tilde \epsilon
_s$ may be very close. Since in the above expressions (\ref{eqs}) the
non-diagonal matrix elements of the interaction are not taken into account,
one can expect that the distribution of occupation numbers defined by these
equations gives a correct result if the interaction is weak enough (ideal
gas approximation). However, we can show (see Section 8) that, in fact, even
for strong interaction the Fermi-Dirac distribution can be also valid if the
total energy $E$ is corrected in a proper way, by taking into account the
increase of the temperature due to statistical effects of interaction.

One should also note that somewhat similar procedure transforms the
canonical distribution (\ref{gibbs}) into the Fermi-Dirac distribution (see
e.g. \cite{book}) in the case of many non-interacting particles (ideal gas).
It is curious that the Fermi--Dirac distribution is very close to the
canonical distribution (\ref{gibbs}) even for very small number of particles
($n=2$), provided the number of essentially occupied orbitals is large
(which happens for $T\gg \epsilon $ or $\mu \gg \epsilon $ ). In fact, this
is a result of a large number of ``principal'' terms in the partition
function $Z_s\,$ which allows us to replace $\alpha _s$ by $\alpha $ in the
term $Z_s(n,T)/Z_s(n-1,T)\equiv \exp (\alpha _s+\beta T)$ in the canonical
distribution (\ref{gibbs}) (compare with (\ref{nsZ})).

More accurate consideration shows, however, that the temperature $T\,$ in
the Fermi-Dirac distribution is different from that in the canonical
distribution. Indeed, using the expansion $\alpha _s=\alpha (\epsilon
_F)+\alpha ^{\prime }(\epsilon _s-\epsilon _F)$ one can obtain the relation
between the Fermi-Dirac ($\beta _{FD})$ and canonical ($\beta )$ inverse
temperatures, $\beta _{FD}=\beta +\alpha ^{\prime }\epsilon _F$ . Concerning
the chemical potential, its definition also changes, $-\mu /T=\alpha
(\epsilon _F)-\alpha ^{\prime }\epsilon _F$ . This fact is confirmed by our
numerical simulations for an isolated system with few interacting
Fermi-particles \cite{FIC96,FGI96,FI97}. Namely, for the same total energy $%
E $ of the system , the canonical and Fermi-Dirac distributions give the
same distribution $n_s$ defined, however, by different temperatures since
they are determined by different equations (\ref{energy}) and (\ref{eqs}).

The closeness of these two distributions for any number of particles is not
so surprising in the presence of the thermostat, where even one particle is
in the equilibrium. To the contrary, for isolated systems with small number
of particles the applicability of the Fermi-Dirac distribution is not
obvious. To answer this question, one needs to analyze the role of
interaction in the creation of an equilibrium distribution.

\section{Analytical calculation of occupation numbers in finite systems}

The advantage of the approach developed in this paper is that if we know the
shape of eigenstates in the many-particle basis of non-interacting particles
(the $F-$ function) and the unperturbed density of states $\rho _0(E)$, one
can analytically calculate the distribution of the occupation numbers $n_s$.

In order to calculate the occupation numbers $n_s$, we use the expression (%
\ref{nsZ}) containing two partial partition functions $Z_s(n,E)$ and $%
Z_s(n-1,E-\epsilon _s)$ which correspond to systems with $n$ and $n-1$
particles, with the orbital $s\,$ is excluded from the set of
single-particle states. The partition function can be found from the
relation 
\begin{equation}
\label{Z}Z=\sum_kF(E_k-E)\approx \int \rho _0(E_k)F(E_k-E)dE_k 
\end{equation}
The density of unperturbed states $\rho _0(E_k)$ in a system of $n$
particles distributed over $m$ single-particle states (orbitals) was shown
to be close to the Gaussian (see for example,\cite{FW70,BF71,brody}): 
\begin{equation}
\label{rho0}\rho _0(E_k)=\frac N{\sqrt{2\pi \sigma _0^2}}\exp \left( -\frac{%
\left( E_k-E_c\right) ^2}{2\sigma _0^2}\right) 
\end{equation}
with $E_c$ as the center of the energy spectrum and $N$ as the total number
of states. Let us assume that the shape of eigenstates $F-$ is given by the
Gaussian, too: 
\begin{equation}
\label{Fgauss}F(E_k-E)=\frac 1{\sqrt{2\pi (\Delta E)^2}}\exp \left( -\frac{%
\left( E_k-E\right) ^2}{2(\Delta E)^2}\right) 
\end{equation}
Here the variance $(\Delta E)^2$ is defined by Eqs.(\ref{delE},\ref{50}),
and $E=E^{(i)}+\Delta _1^{(i)}=H_{ii}$ (see Appendix 1).

As was found, Gaussian shape of eigenstates (apart from long tails) occurs
in realistic systems like Ce atom \cite{FGGK94} and heavy nuclei \cite{ZEL97}
. Recently, the form of the $F-$ function in dependence on the perturbation
has been studied in details \cite{CFI97} in the model of Wigner Band Random
Matrices , as well as in the random two-body interaction model \cite{FI97} 
(see also nuclear calculations \cite{Lew94,Lau95}). 
In particular, it was discovered, that the gaussian-type shape happens when
the interaction is large enough, namely, when the Breit-Wigner width $\Gamma
=2\pi \rho V^2$ is comparable with the root-mean-square width $\Delta E$ (an
effective band width of a Hamiltonian matrix).

By performing the integration in Eq.(\ref{Z}) one gets 
\begin{equation}
\label{ZE}Z(E)=\frac N {\sqrt{2\pi \sigma ^2}}\exp \left( -\frac{\left(
E-E_c\right) ^2}{2\sigma ^2}\right) 
\end{equation}
where $\sigma ^2=\sigma _0^2+(\Delta E)^2$ (it coincides with the variance
of the perturbed spectrum). In order to calculate the occupation numbers $n_s
$, we use the expression (\ref{nsZ}). For this, one needs to find the
partial partition functions $Z_s(n,E)$ and $Z_s(n-1,E-\epsilon _s)$
corresponding to $n$ and $n-1$ particles with the orbital $s\,$ is excluded
from the set of single-particle states. Now we have to calculate the number
of states $N_s$ and the center $E_{cs}$ for these truncated systems,%
$$
N_s(n,m-1)=\frac{(m-1)!}{(m-1-n)!\,n!} 
$$
$$
\,N_s(n-1,m-1)=\frac{(m-1)!}{(m-n)!\,(n-1)!} 
$$
$$
E_{cs}(n)=\overline{\epsilon _{-s}}\,n\,;\,\,\,\,E_{cs}(n-1)=(\overline{%
\epsilon _{-s}})(\,n-1) 
$$
$$
\,\,\overline{\epsilon _{-s}}=\frac{\sum_{p\neq s}\epsilon _p}{m-1} 
$$
The variance $\sigma _{0s}$ of the energy distributions can be estimated as%
$$
\sigma _{0s}^2(n)\approx \sigma _{1s}^2\,n\,\,\,;\,\,\,\sigma
_{0s}^2(n-1)\approx (\sigma _{1s}^2)\,(n\,-1)\,\,\, 
$$
where $\sigma _{1s}^2$ is the variance of single-particle spectrum with the
excluded orbital $s$ . Here, for simplicity, we have neglected the Pauli
principle which is valid for $m\gg n$ . More accurate calculation can be
easily done with the use of calculator. As a result, the distribution of
occupation numbers has the form%
$$
\overline{n_s}(E)=\frac 1{1+R} 
$$
$$
R=\frac{m-n}n\frac{\sigma _s(n-1)}{\sigma _s(n)} 
$$
\begin{equation}
\label{R}\times \exp \left[ -\frac{\left( E-E_{cs}(n)\right) ^2}{2\sigma
_s^2(n)}+\frac{\left( E-\epsilon _s-E_{cs}(n-1)\right) ^2}{2\sigma _s^2(n-1)}%
\right] 
\end{equation}
where $\sigma _s^2=\sigma _{s0}^2+(\Delta E)^2$ .

It is instructive to compare this result with the Fermi-Dirac distribution
which is valid for large number of particles. In this case $R=\exp
((\epsilon _s-\mu )/T_{th})$ where $T_{th}=\sigma ^2/(E_c-E)$ is the
thermodynamic temperature which is discussed below, see (\ref{Ttherm}). The
chemical potential $\mu $ in this case should be calculated numerically to
fix the total number of particles $n$. The data are reported in Fig.1. One
can see that Eq.(\ref{R}) predicts occupation numbers in perfect agreement
with the numerical experiment.

Finally, note that the same method can be used to solve another problem: to
find the distribution of the occupation numbers $n_s(T)$ in finite systems
of interacting particles in the thermal bath with the temperature $T$ . For
such a case, it is enough to replace the $F-$function by the canonical
average $F(T,E_k)$, see (\ref{FT}). In fact, it is the method for taking
into account the ``random'' interaction in the canonical distribution. The
result for the occupation numbers $n_s(T)$ can be obtained from (\ref{R}) by
replacing $\sigma _s^2\rightarrow 2\sigma _s^2$ and $E\rightarrow E_m+\Delta
_1(E_m)$ where $E_m=E_c-\sigma ^2/T$ and $\Delta _1$ is a small correction,
see Appendix 1.

\section{Particles and quasi-particles, role of single-particle spreading
width}

In previous sections we have discussed the distribution of occupation
numbers for real particles distributed over given orbitals. At the same
time, there exist traditional approach which is based on the notion of
``quasi-particles''. It allows to incorporate the effects of interaction in
terms of single-particle states and goes beyond the mean-field
approximation. As is well known, the interaction leads to the spreading
width $\gamma _s$ for single-particle orbitals . It also results in the
shift of average energies, $\tilde \epsilon _s=\epsilon _s+\delta \epsilon
_s $ . According to our numerical data (\cite{FIC96}) for the random
two-body interaction model, the shifts $\delta \epsilon _s$ turn out to be,
in average, smaller than $\gamma _s$ and for this reason one can take into
account the effect of spreading widths $\gamma _s$ only.

Here we would like to analyze the role of the spreading widths for the
distribution of occupation numbers and compare with our approach where the
interaction is taken into account in terms of many-body states. For this,
let us average the standard Fermi-Dirac occupation numbers $n_s$ over the
energy interval $\gamma _s$: 
\begin{equation}
\label{ngamma}n_s=\int\limits_{\epsilon _s-\gamma _s/2}^{\epsilon _s+\gamma
_s/2}n(\epsilon )\frac{d\epsilon }{\gamma _s}=1-\frac T{\gamma _s}\ln \left[ 
\frac{1+\exp \frac{(\epsilon _s+\frac{\gamma _s}2-\mu )}{2T}}{1+\exp \frac{%
(\epsilon _s-\frac{\gamma _s}2-\mu )}{2T}}\right] 
\end{equation}
where 
\begin{equation}
\label{nFD}n(\epsilon )=\frac 1{1+\exp (\frac{\epsilon -\mu }T)} 
\end{equation}
It seems, this is the simplest form of the Fermi-Dirac distribution for the
``quasi-particles'' with finite spreading widths. One can check that in the
limit $\gamma _s=0$ the Fermi-Dirac expression (\ref{nFD}) with $%
n_s=n(\epsilon _s)$ is recovered.

To test the sensitivity of the occupation numbers to the values of the
spreading widths $\gamma _s$, we have solved equations (\ref{eqs}) for
chemical potential and temperature (for a given energy of our isolated
system) using the standard expression (\ref{nFD}), and compared the result
with that obtained by using the expression (\ref{ngamma}), see details in 
\cite{FIC96}. The data have revealed that the chemical potential practically
does not change, while there is a noticeable decrease of the temperature, $%
T(\gamma \neq 0)<$ $T(\gamma =0)$ . The striking result is that the two
curves for the occupation numbers (\ref{ngamma}) and (\ref{nFD}) coincide
with a high accuracy, namely, $n_s(\epsilon _s,\gamma ,T)\approx
n_s(\epsilon _s,\gamma =0,T+\Delta T)\equiv n_s(\epsilon _s,\tilde T).$ This
means that the temperature mimics the effect of the spreading widths, the
phenomenon which is far from being trivial. The shift of the temperature for 
$\gamma \ll \mu \,$ can be estimated analytically as $\Delta T\approx \gamma
^2/16T.\,$

The above result indicates that one should not worry about the finite width
of single-particle orbitals (``quasi-particles'') when calculating the
occupation numbers.

\section{Thermodynamics of small systems}

One of the important questions is about thermodynamical description of
isolated systems of interacting particles. In any thermodynamical approach
one needs to define, in a consistent way, such quantities as entropy,
temperature and equation of state. Different definitions of the entropy and
temperature have been recently discussed \cite{zel2} (see also \cite{zel93})
 in application to shell
models of heavy nuclei. In particular, it was found that for a realistic
residual interaction, different definitions of temperatures lead to the same
result. Below, we analyze few definitions of the temperature and entropy and
show that for small number of interacting particles they may give quite
different results.

Standard thermodynamical definitions of the entropy $S_{th}\,$and
temperature are based on the density of states $\rho (E)$, 
\begin{equation}
\label{St}S_{th}=\ln \,\rho (E)\,+\,const 
\end{equation}

\begin{equation}
\label{Ttnew}\frac 1{T_{th}}=\frac{dS_{th}}{dE}=\frac{d\ln \,\rho }{dE} 
\end{equation}
In fact, such a definition of the temperature follows from the estimate of
the position of maximum of the canonical averaging function $\Phi _T(E)$ ,
see (\ref{fi}) and (\ref{Tt}). It is usually assumed that the position of
its maximum $E_m$ coincides with the energy $E$ of a system. One should
stress that in the above definitions $\rho (E)$ is the perturbed density of
states, therefore, the interaction is essentially taken into account.

However, for finite isolated systems with the fixed energy $E$ , the
definition of the temperature given by the relation $\left\langle
E\right\rangle _T=E$ (see Eq.(\ref{energy}) seems to be more natural. Here
the averaging is performed over the canonical distribution (\ref{gibbs}).
Since the width $\Delta _T$ of the canonical averaging function $\Phi _T(E)$
is not zero, the two definitions of the temperature, (\ref{Ttnew}) and (\ref
{energy}) give, in principal, different results. Indeed, in the case of the
Gaussian form of $\rho (E)$ the value of $T_{th}$ given by (\ref{Ttnew})
takes the form (see also \cite{zel2}), 
\begin{equation}
\label{Ttherm}T_{th}=\frac{\sigma ^2}{E_c-E} 
\end{equation}
where $E_c$ and $\sigma $ are the center and the width of the distribution $%
\rho (E).$

On the other hand, direct evaluation of the relation (\ref{energy}) leads to
the following definition of the temperature, 
\begin{equation}
\label{Tcan}T_{can}=\frac{\sigma ^2}{E_c-E+\Delta }
\end{equation}
Here, the shift $\Delta $ is given by the expression 
\begin{equation}
\label{DeltaT}\Delta =\frac \sigma K\left[ e^{-\frac{\left( E_{\min
}-E_m\right) ^2}{2\sigma ^2}}-e^{-\frac{\left( E_{\max }-E_m\right) ^2}{%
2\sigma ^2}}\right] 
\end{equation}
where%
$$
K=\int\limits_{x_{\min }}^{x_{\max }}\exp \left( -\frac{x^2}2\right)
dx\approx \sqrt{2\pi }\,\,; 
$$
\begin{equation}
\label{K}\,\,\,\,x=\frac{E-E_m}\sigma \,\,;\,\,\,E_m=E_c-\frac{\sigma ^2}{%
T_{can}}
\end{equation}
One can see that the shift $\Delta $ itself depends on the temperature and
is proportional to the width $\Delta _T=$$\sigma $ of the function $\Phi
_T(E)$. In the above relations, $E_{\min }$ and $E_{\max }$ are the low and
upper borders of the energy spectrum. Note that the relation $\Delta =0$
occurs at the center of the spectrum, therefore, the temperature in the
upper part of the spectrum is negative (it is typical for systems with
bounded spectrum, for example, for spin systems). In fact, our model (\ref{H}%
) with finite number $m$ of orbitals can be treated as the model of one open
shell in atoms, nuclei, clusters, etc. However, in realistic many-body
systems there are always higher shells which contribute to the density of
states for higher energy. Thus, the density of states $\rho (E)$ is a
monotonic function which results in the positive temperature. For such
physical applications, the model (\ref{H}) with finite number of orbitals is
reasonable in the lower part of the energy spectrum where the influence of
higher shells can be neglected.

One can also see that the difference between the two equations of state $%
T(E) $ defined by Eqs.(\ref{Ttherm}) and (\ref{Tcan}), disappears for highly
excited eigenstates (for which $E_m-E_{\min }\gg \sigma ),$ or in large
systems with $n\gg 1$ . Indeed, one can obtain, $E_c-E\sim n\sigma _1$ ,
where $\sigma _1$ is the width of single-particle spectrum. On the other
hand, according to the central limit theorem, the variance of total energy
spectrum can be estimated as $\sigma _0^2\approx \sum_n\sigma _1^2=n\sigma
_1^2$ , therefore, the ratio $\sigma /(E_c-E)\sim 1/\sqrt{n}$ tends to zero
at $n\rightarrow \infty $ . Note, that in finite systems (atom, nucleus
etc.) the number of valence particles (particles in an open shell) is not
large. For example, for Ce atom we have $n=4$ \cite{FGGK94} and in nuclear
shell model \cite{ZELE} $n=12$ , therefore, the corrections to the
thermodynamical temperature (\ref{Ttnew}) can be significant, especially,
for low energies. Here, we do not take into account particles from deep
closed shells since their excitation energy is high and they do not
contribute to the thermodynamical and statistical properties of systems
(though they renormalize parameters of the Hamiltonian (\ref{H}) describing
the interaction between valence particles).

The energy dependence of temperatures $T_{th}$ and $T_{can}$ is shown in
Fig.2. The data are given for the model of $n=4$ interacting
Fermi--particles distributed over $m=11$ orbitals. The two-body interaction
is taken to be completely random, given by the gaussian distribution of
two-body matrix elements with $V=0.12$ ; this value should be compared with
the mean energy distance $d_0=1$ between the orbitals (single-particle
energies), see details in \cite{FIC96,FGI96,FI97}. The comparison of the
thermodynamical temperature $T_{th}$ defined by (\ref{Ttherm}) with the
``canonical'' temperature (\ref{Tcan}) reveals quite strong difference in
all the range of the rescaled energy $\chi =(E-E_{fermi})/(E_c-E_{fermi})$.
To test our analytical expression for the canonical temperature $T_{can}\,$%
we have performed direct numerical calculation of the temperature according
to (\ref{energy}) with the actual spectrum $E^{(i)}\,$of the two-body random
interaction (instead of the Gaussian approximation of $\rho (E)$ ).
Numerical results well agree with the analytical expression (\ref{Tcan}).

The knowledge of the equation of state $E(T)$ gives the possibility to
examine heat capacity of closed systems with finite number of interacting
particles, 
\begin{equation}
\label{heat}C=\frac{dE}{dT}=\frac{\sigma ^2}{T^2}\left( 1+\frac{\partial
\Delta }{\partial E_m}\right) 
\end{equation}
The second term in the above expression is a correction which vanishes for
highly excited states or in large systems, however, it may be important in
other situations.

Following the paper \cite{zel2}, we can also compare different definitions
of entropy. Natural definition of the entropy in isolated systems can be
directly related to the number of principal components $N_{pc}$ in exact
eigenfunctions., 
\begin{equation}
\label{Sef}S_{EF}=\ln \,N_{pc}
\end{equation}
In such a definition the entropy characterizes the complexity of a system
(note, that for unperturbed ``simple'' states $N_{pc}=1$ and $S_{EF}=0$ ).
There are several definitions of $N_{pc}$ one of which is the so-called
``entropy localization length'' defined via the information entropy $S_{\inf
}$ of eigenstates, 
\begin{equation}
\label{Npcinf}N_{pc}=\exp (S_{\inf })
\end{equation}
where%
$$
S_{\inf }(E)=-\sum_kF_k(E)\ln F_k(E) 
$$
\begin{equation}
\label{SinfF}\approx \int dE_k\rho (E_k)F(E_k-E)\ln F(E_k-E)
\end{equation}
Here we have used the $F-$function instead of $\left| C_k^{(i)}\right| ^2$
in order to have smooth dependence of the entropy $S_{\inf }$ on the energy $%
E$ . Another possibility is to find $N_{pc}$ from the ``inverse
participation ratio'', 
\begin{equation}
\label{NpcIPR}N_{pc}^{-1}\,=\,\sum_k\left( F_k(E)\right) ^2
\end{equation}
One more definition is $N_{pc}^{-1}=\max [F_k(E)]\approx F(E_k=E)$ which has
been used in \cite{FGGK94}. The difference between the above definitions of $%
N_{pc}$ depends on a specific shape of $F_k(E)$ , however, the values of $%
N_{pc}$ differ from each other by some coefficient which is typically close
to 1.

On the other hand, the estimate for $N_{pc}$ can be obtained simply from the
relation 
\begin{equation}
\label{Ngamma}N_{pc}\approx \frac \Gamma D=\Gamma \rho (E) 
\end{equation}
where $\Gamma $ is the spreading width of the function $F_k(E)$ and $D$ is
the local mean spacing between many-particle energy levels. Thus, one can
directly relate the number of principal components $N_{pc}$ to the density
of states $\rho (E)$ , 
\begin{equation}
\label{SefGamma}S_{EF}=\ln \,N_{pc}\,\approx \,\ln \,\rho (E)\,+\ln \,\Gamma 
\end{equation}
One can see that the entropy $S_{EF}$ found from exact eigenstates coincides
with the thermodynamical entropy $S_{th}$ if the second term in ($\ref
{SefGamma})\,$ does not depend on the energy. As is shown in Appendices 1-3,
the spreading width $\Gamma \,$ only weakly depends on the energy, in
contrast to a very strong energy dependence of $\rho (E)$ . The fact that
the information entropy $S_{\inf }$ contains the term $\ln \,\rho $ was
mentioned for the first time in \cite{zel2}. One should stress that the
above relations ($\ref{Ngamma},\ref{SefGamma})$ are valid if $N_{pc}$ is
smaller than the size $N$ of many-particle basis, $N_{pc}<N/2$ . One has
also to remind that systems under consideration are assumed to be in
equilibrium, see discussion in Section 9.

\section{Increase of effective temperature due to statistical effects of
interaction}

In Section 4 we have shown that in the case of large number of particles,
the distribution of occupation numbers is of the Fermi-Dirac form (\ref
{FDtype}) if the local mean field approximation is valid. However, if one
uses the expression (\ref{eqs}) in order to find the chemical potential $\mu 
$ and temperature $T$ , one can obtain inaccurate result. To demonstrate
this, we have computed the distribution of occupation numbers $n_s$ for the
two-body random interaction model directly from exact eigenstates of the
Hamiltonian matrix (\ref{H}) defined in the basis of many-particle
unperturbed states (see also \cite{FIC96,FI97}). These data for the
``experimental'' values of $n_s$ are shown in Fig.3 by the histogram which
is obtained by the average over a small energy window in order to smooth the
fluctuations (also, an additional averaging over different realizations of
the random two-body random interaction has been done). To compare with the
standard Fermi-Dirac distribution, we have numerically solved Eqs.(\ref{eqs}%
) in order to find the temperature and chemical potential. The resulting
distribution of the occupation numbers $n_s$ is shown in Fig.3 by circles.
One should stress that the value of the energy $E$ in (\ref{eqs}) was taken
the same as for the exact eigenstates from which actual distribution of $n_s$
was computed, namely, $E\approx E^{(i)}\,$. The comparison of the actual
distribution (histogram) with the theoretical one, see (\ref{eqs}), reveals
big difference for a chosen (quite strong) perturbation $V=0.20\,.$

This discrepancy is due to the fact that the off-diagonal interaction is not
taken into account in (\ref{eqs}). In \cite{FIC96} it was pointed out that
one can take into account statistical effects of interaction and to correct
the total energy in (\ref{eqs}) in the following way. Let us first consider
the thermodynamical temperature $T_{th}$ defined by (\ref{Ttherm}). As was
mentioned above, see (\ref{sigma}), one can represent the width of perturbed
density of states $\rho (E)$ in the form $\sigma ^2=\sigma _0^2+ \overline{%
(\Delta E)^2}$ where $\sigma _0$ relates to the unperturbed density and $%
\Delta E$ is an increase of the width of the energy spectrum due to the
interaction. Thus, the temperature at a given energy $E$ increases due to
the interaction as follows 
\begin{equation}
\label{TdT}T=T_0+\Delta T\approx \frac{\sigma _0^2}{E_c-E}+\frac{\overline{%
(\Delta E)^2} }{E_c-E} 
\end{equation}
resulting in the following relation, 
\begin{equation}
\label{ddt}\Delta T/T_0=\overline{(\Delta E)^2}/\sigma _0^2 
\end{equation}

The explanation of this increase of the temperature was given in \cite{FI97}
and reads as follows. Since the density of states rapidly increases with the
energy $E$ , the number of higher basis states admixed to an eigenstate by
the interaction is larger than the number of lower basis states. An extreme
example is the ground eigenstate, which contains basis components of higher
energies only. As a result, the mean energy 
\begin{equation}
\label{Eki}\left\langle E_k\right\rangle _i=\sum_kE_kF_k^{(i)}\approx \int
E_kF_k^{(i)}\rho _0(E_k)dE_k
\end{equation}
of the components in an exact eigenstate $\left| i\right\rangle $ is higher
than the eigenvalue $E^{(i)}$ corresponding to this eigenstate (we consider
here eigenstates in the lower part of the spectrum only). There is another
effect which gives the increase of $\left\langle E_k\right\rangle _i-E^{(i)}$
even if the density of states does not depend on the energy. Due to
repulsion between the energy levels, the eigenvalues move down for this part
of the spectrum, therefore, the difference between $\left\langle
E_k\right\rangle _i$ and $E^{(i)}$ increases due to the interaction. This
second effect shifts the ``center'' of the function $F_k^{(i)}=F(E_k-E^{(i)})
$ . One should stress that all effects leading to the above shift of the
energy are automatically taken into account in the relation (\ref{Eki}).
Thus, one can analytically calculate this shift $\Delta _E=\left\langle
E_k\right\rangle _i-E^{(i)}$ from the equation (\ref{Eki}). For this, one
needs to know the unperturbed density of states and the form of the $F-$%
function. The evaluation of the shift $\Delta _E$ has been done in \cite
{FI97} by assuming some form for the $F-$ function which is valid in a wide
range of the interaction strength $V$ ,%
$$
\Delta _E=\left\langle E_k\right\rangle _i-E^{(i)}=\frac{d\ln \rho _0}{dE}
\overline{\left( \Delta E\right) ^2}=\frac 1{T_0}\overline{\left( \Delta
E\right) ^2} 
$$

\begin{equation}
\label{Delfinal}=\frac{\overline{\left( \Delta E\right) ^2}}{\sigma _0^2}%
\left( E_c-E\right) 
\end{equation}

Now, in order to obtain the corresponding increase of the temperature, one
should insert the shifted energy $E\equiv \left\langle E_k\right\rangle
_i=E^{(i)}+\Delta _E$ into the equation for the temperature,%
$$
T=T_0+\Delta T=\frac{\sigma _0^2}{E_c-E^{(i)}-\Delta _E} 
$$
\begin{equation}
\label{Tfinal}\approx \frac{\sigma _0^2}{E_c-E^{(i)}}+\frac{\overline{\left(
\Delta E\right) ^2}}{E_c-E^{(i)}}
\end{equation}
One can see that Eq.(\ref{Tfinal}) is consistent with Eq.(\ref{TdT}) for $
\overline{\left( \Delta E\right) ^2}\ll \sigma _0^2$ .

Thus, to find correct values for the occupation numbers in the Fermi-Dirac
distribution, we should substitute the increased energy $E=E^{(i)}+\Delta
_E=\left\langle E_k\right\rangle _i$ into equations (\ref{eqs}) for the
chemical potential and temperature. The resulting shift of the temperature
and chemical potential leads to the distribution of the occupation numbers
shown in Fig.3 by diamonds. As one can see, such a correction gives a quite
good correspondence to the numerical data.

To check the analytical prediction (\ref{Delfinal}) for the shift $\Delta _E$
, we have calculated this shift directly by comparing the energy $E^{(i)}$
of exact eigenstates with the energy $\left\langle E_k\right\rangle _i$ .
The latter has been computed from the exact relation $\left\langle
E_k\right\rangle _i=\sum_kE_k\left| C_k^{(i)}\right| ^2$ (compare with (\ref
{Eki})). The comparison of these data (circles in Fig.4) with Eq. (\ref
{Delfinal}) (straight full line) shows a good agreement, if to neglect
strong fluctuations around the global dependence. These fluctuations are due
to fluctuations in the components of specific exact eigenstates $\left|
i\right\rangle $ (note, that the presented data correspond to the individual
eigenstates, without any additional averaging).

Finally, we would like to note that the described above method can be also
used to solve the ``canonical'' problem of finding the distribution of the
occupation numbers for a system of interacting particles in the thermal
bath. Indeed, the distribution $n_s(T)$ for the ``ideal'' gas is given by
the canonical distribution, or more simple, by the Fermi-Dirac distribution
if the number of particles $n\gg 1$ (in practice, $n>2$ is enough). The
increase of ``kinetic'' energy due to the random residual interaction is
given by $\Delta _E=\overline{(\Delta E)^2}/T$ , see Eq.(\ref{Delfinal}).
Then, one can find the effective temperature which corresponds to the same
increase of the average ``kinetic'' energy in the ideal gas. It can be done
by the differentiation of $T^{-1}=dS/dE$ , where $S$ is the entropy, $S=\ln
\,\rho _0+const$ . The result reads as 
\begin{equation}
\label{deltaEnew}\Delta T=-T\frac{d^2\ln \rho _0}{dE^2}\overline{(\Delta E)^2%
} 
\end{equation}
For the Gaussian shape of the level density $\rho _0(E)$ it coincides with (%
\ref{ddt}). Thus, we can use the Fermi-Dirac distribution with the effective
temperature $T_{eff}=T+\Delta T$ in order to describe ``randomly''
interacting particles (for $\Delta T\ll T$) in the heat bath.

\section{Criteria for the onset of chaos, equilibrium and thermalization}

The theory presented above is based on the notion of chaos in terms of
statistical properties of compound eigenstates. Typically, the onset of
quantum chaos is associated with large number of components in the
eigenstates. However, as we can see below, this condition is not enough for
the emergence of equilibrium distributions and, in essence, there are
different regimes of ``chaos''. Here, we analyze the conditions under which
the possibility of statistical description of isolated systems of
interacting particles can be directly related to statistical properties of
eigenstate components.

Let us start with the condition of a large number of principal components $%
N_{pc}$ provided there is an equilibrium distribution of the components of
eigenstates. For the relatively small interaction, the distribution of the
components has the Breit-Wigner form, 
\begin{equation}
\label{BWform}F_k^{(i)}\equiv \overline{\left| C_k^{(i)}\right| ^2}=\frac
1{N_{pc}}\frac{\Gamma _{BW}^2/4}{\left( E_k-E^{(i)}\right) ^2+\Gamma
_{BW}^2/4} 
\end{equation}
Here, the value $N_{pc}=\frac{\pi \Gamma _{BW}}{2D}$ is defined by the
normalization condition and $D=\rho ^{-1}(E)$ is the mean spacing between
energy levels, and spreading width is given by 
\begin{equation}
\label{GammaDf}\Gamma _{BW}=2\pi \frac{V^2}{d_f} 
\end{equation}
where $V^2$ is the mean squared value of matrix elements of the two-body
interaction, see (\ref{H}). In the denominator $d_f$ stands for the mean
energy spacing of those basis components to which a particular basis state $%
\left| k\right\rangle $ can ``decay'' due to direct two-body interaction,
see details in Appendix 3. Thus, the condition $N_{pc}\gg 1$ reads%
$$
N_{pc}=\frac{\pi ^2V^2}{d_fD}\gg 1 
$$
or 
\begin{equation}
\label{crit}V\gg \frac 1{\pi ^2}\sqrt{d_fD} 
\end{equation}

However, as was pointed out in \cite{AGKL97} (see also 
\cite{SS97,MF97,JS97}), there is a phenomenon of 
``localization in the Fock space''
which means that the states with different number of excited particles do
not mix with each other. This situation occurs when $V<d_f$ (with some
logarithmic corrections). This means that in order to have an ergodic
distribution for the eigenstate coefficients resulting in the equilibrium
distribution of occupation numbers for individual eigenstates, one needs
both the conditions, $V>d_f$ and (\ref{crit}). Since $d_f$ is typically much
bigger than $D$ , the condition $V> d_f$ is stronger than (\ref{crit}).

Let us now discuss the properties of eigenstates and distribution of
occupation numbers in dependence on two above parameters, $V/d_f$ and $%
N_{pc} $ . Since the value of $N_{pc}$ increases with an increase of $V$, we
first start with a very weak interaction for which exact eigenstates have
only a few relatively large components ($N_{pc}\sim 1)$. In such a case the
eigenstates are strongly localized in the unperturbed basis and, therefore,
can be described by conventional perturbation theory. This situation is
quite typical for lowest eigenstates (where the density of states is small)
even if for higher energies the eigenstates can be considered as very
``chaotic'' ($N_{pc}\gg 1)$ . We term this region (I) as a region of strong
(perturbative) localization.

The second region (II) is characterized by an ``initial chaotization'' of
exact eigenstates and corresponds to a relatively large, $N_{pc}\gg 1$
number of principal components and $V<d_f$. The latter condition is
essential since it results in very strong (non-Gaussian) fluctuations of
components $C_k^{(i)}$ \cite{FGS97} for the fixed energy $E^{(i)}$ of
compound state $\left| i\right\rangle \,.$ Such a type of fluctuations
reflects itself in a specific character of eigenstates, namely, they turn
out to be ``sparsed''. As a result, the number of principal components can
not be estimated as $N_{pc}\approx \Gamma /D$ , as is typically assumed in
the literature. Let us note that the energy width $\Gamma $ of eigenstates
is still close to the expression (\ref{GammaDf}).

The above specific properties of compound states can be explained by the
perturbation theory in the parameter $V/d_f$ . Indeed, in the zero order
approximation an eigenstate $\left| i\right\rangle $ coincides with a basis
state $\left| k_0\right\rangle $ for which the particles occupy definite
orbitals. In the first order in $V/d_f$ the eigenstate $\left|
i\right\rangle $ is constructed by those basis states $\left|
k_1\right\rangle $ which can be obtained from $\left| k_0\right\rangle $ by
moving one or two particles (due to the two-body character of the
interaction). As a result, the coefficients of the state $\left|
i\right\rangle $ can be estimated as $%
C_{k_1}^{(i)}=V_{k_{0,}k_1}/(E^{(i)}-E_{k_1})$ . If the matrix elements $%
V_{k_{0,}k_1}$ are Gaussian variables, for the fixed spacing $\left|
E^{(i)}-E_{k_1}\right| $ the coefficients $C_k^{(i)}$ are also distributed
according to the Gaussian.

The situation is completely different in higher orders. For example, in the
second order we have $C\equiv
C_{k_2}^{(i)}=%
\sum_{k_1}V_{k_{0,}k_1}V_{k_{1,}k_2}/((E^{(i)}-E_{k_1})(E^{(i)}-E_{k_2}))\,%
\, $ and the distribution $P(C)\,$ (for the fixed $\left|
E^{(i)}-E_{k_2}\right| $ ) is close to the Lorentzian, namely, $P(C)\sim
C_{cr}/C^2$ for $C_{cr}<C<1$ where $C_{cr}\sim V^2/d_f^2\,$ \cite{FG94,FGS97}%
. Long tails in the distribution $P(C)$ are due to the possibility of small
values of the denominator $E^{(i)}-E_{k_1}$ in intermediate states. The
higher orders $r$ of the perturbation theory give stronger fluctuations
(with additional logarithm in $P(C)\,)$ because of large number of small
denominators. Another type of even stronger fluctuations result from very
different nature of many-body basis states with close energies. Namely, the
states with nearly the same energy can differ by a large number of moved
particles in order to obtain the corresponding states. Therefore, the
transition into close (in energies) states appear in different orders of the
perturbation. Therefore, the fluctuations of the components $C$ appear to be
abnormal, without obeying the standard central limit theorem. In particular,
the above effect leads to very large fluctuations in the distribution of the
occupation numbers $n_s$, as a function of the energy $E$ of compound
states. Specifically, the fluctuations do not decrease as $N_{pc}^{-1/2}$ ,
for example, if $P(C)\sim 1/C^2$ the fluctuations of $n_s(E)$ do not depend
on $N_{pc}$ at all for $N_{pc}<1/C_{cr}\,$(see, for example, \cite{FG94}).

As an example, one can take the two-body random interaction model (\ref{H})
with $n=4$ particles and $m=11$ orbitals and very weak perturbation $%
V/d_0\approx 0.02$ where $d_0$ is the mean level spacing between
single-particle energy levels, see Fig.5. One can see that the
``experimental'' distribution of occupation numbers has nothing to do with
the Fermi-Dirac distribution (full diamonds), it turns out to be even the
non-monotonic function of the energy $\epsilon _s$ of orbitals (see also 
\cite{FIC96}). Note that the averaging procedure used in Fig.5 has not wash
out the fluctuations in $n_s$ .

With further increase of the interaction, where $N_{pc}\gg 1$ and $V>d_f\,$
, the region (III) of the equilibrium for the $F-$distribution (\ref{nalpha}%
) emerges. In this region the fluctuations of the eigenstate components $%
C_k^{(i)}$ are of the Gaussian form \cite{FGS97} which leads to small
fluctuations of the occupations numbers $n_s$ in accordance with the central
limit theorem for the sum (\ref{ns}), $\Delta n_s/n_s\sim N_{pc}^{-1/2}\ll 1$
for $n_s\sim 1$ . One should stress that in this region the value of $N_{pc}$
is given by the common estimate, $N_{pc}\sim \Gamma /D$ . As a result, the
distribution of occupation numbers changes slightly when changing the energy
of a system. Such a situation can be naturally associated with the onset of
thermal equilibrium, though the form of the distribution $n_s$ can be quite
different from the Fermi-Dirac distribution. In this case, the $F-$%
distribution (\ref{nalpha}) gives a correct description of an actual
distribution of occupation numbers in isolated quantum systems of
interacting particles. One can see that the equilibrium distribution for the
occupation numbers arises for much weaker condition compared to that needed
for the Fermi-Dirac distribution. Since the energy interval $d_f$ between
two-particle-one-hole energy levels is small, it is enough to have a
relatively weak residual interaction $V>d_f$ in order to have the
equilibrium distribution (note, that the value of $d_f$ decreases rapidly
with the excitation energy, see Appendix 3).

Next region (IV) is where the canonical distribution (\ref{gibbs}) occurs;
for this case in addition to the equilibrium, one needs to have large number
of particles, $n\gg 1.$ If, also, the condition $\Gamma \ll nd_0$ is
fulfilled, the standard Fermi-Dirac distribution is valid with a proper
shift of the total energy due to the interaction, see Section 8 . Typically,
this region is associated with the onset of the canonical thermalization
(see, for example, \cite{zel2}).

In practice, the condition (IV) of the canonical thermalization is not easy
to satisfy in realistic systems like atoms or nuclei since $n$ in the above
estimates is, in fact, the number of ``active'' particles (number of
particles in a valence shell) rather than the total number of particles.
Thus, the description based on the $F-$distribution (\ref{nalpha}) which
does not require the canonical thermalization condition (IV), is more
accurate.

The above statements are confirmed by the direct numerical study of the
two-body random interaction model \cite{FIC96,FI97} with few particles when
changing the interaction strength $V/d_0.$ If, instead, we increase the
number of particles keeping the interaction small, $V\ll d_0$ , the
distribution (\ref{nalpha}) tends to the Fermi-Dirac one as it is expected
for the ideal gas, see \cite{FI97}.

Finally, we discuss the transition to mesoscopic systems. One can show that
the result strongly depends on the dimensionality $d$ of a system . Let us
consider the case when the number of particles $n$ is fixed, however, the
size $l$ of a system increases. Then, the interval between single particle
energy levels decreases as $d_0\sim l^{-2}$ . Since relative interaction
between two particles decreases like$\,V\sim l^{-d}$ , one can get, $%
V/d_0\sim l^{-(d-2)}$ . Thus, for $d=1$ one has $V\gg d_0$ which means that
strong mixing (chaos) starts just from the ground state. This is in
accordance with the absence of a gap in the distribution of occupation
numbers in 1D case (the so-called Luttinger liquid). To the contrary, in the
3D case we have $V\ll d_0$ ; this means that an admixture of the higher
states to the ground state can be considered perturbatively, which is
consistent with the non-zero gap at $T=0$ . One can see that the transition
between regular region (I) and the equilibrium (region III) in the 3D case
occurs for high states when $V\geq d_f$ . Recently, the question of this
transition has been studied in \cite{AGKL97,MF97,JS97}.

\section{Conclusion}

In this paper we have developed statistical approach to isolated finite
systems of interacting particles, which plays the same role as the canonical
approach for systems in equilibrium with the thermal bath. It can be applied
to complex many-body systems like compound nuclei, rare-earth actinide
atoms, atomic clusters, quantum dots, etc. The key point of this approach is
a new kind of the partition function which is defined to the shape of
compound states ($F-$function) in the many-particle basis of a system
without residual interaction (Slater determinants). It allows to calculate
mean values of different operators as a function of the total energy $E$ of
a system. As an example, we have calculated the occupation numbers $n_s(E)$,
which may be compared with the standard canonical approach giving $n_s(T)$
where $T$ is the temperature of an open system. In large systems
(thermodynamical limit) the distribution of occupation numbers $n_s$ tends
to the canonical distribution with the temperature $T^{-1}=d\,(ln\rho
)\,/\,dE$ where $\rho (E)$ is the energy level density.

Another important area of applications of our approach is the calculation of
non-diagonal matrix elements (transition amplitudes) between the eigenstates
of complex many-body systems. We would like to point out that the approach
can be also used for solving traditional problem of calculations of mean
values of operators in open systems of interacting particles in the
thermostat.

The suggested approach is entirely based on the statistical properties of
chaotic compound states which are due to the two-body interaction between
the particles. For relatively strong interaction the number of components of
compound states is typically large and these components can be treated as
random variables, provided the two-body interaction matrix elements are
``complex'' enough.

The essential question is under which conditions the above approach is valid
in systems with the two-body random interaction. Note, the randomness of the
matrix elements itself is not enough for the onset of the equilibrium in the
system since statistical properties of compound eigenstates essentially
depend on such parameters as the relative strength of the interaction,
excitation energy, number of particles and orbitals (single-particle states
participating in the energy exchange), etc. In particular, even if the
number of principal components in compound states is large, for not strong
enough interaction the statistics of the components can be abnormal, leading
to huge (non-Gaussian) fluctuations in the structure of the eigenstates. In
such a case, there is no equilibrium in the system and standard statistical
description is not valid. In this regime, our numerical data show that the
distribution of the occupation numbers strongly fluctuates when slightly
changing the total energy of the system. Therefore, the transition to the
statistical equilibrium is far from being trivial in the systems with finite
number of interacting particles. However, for larger interaction the
fluctuations of the eigenstate component becomes normal (Gaussian) and the
equilibrium occurs. In this situation the fluctuations of the occupation
numbers $n_s(E)$ are relatively small and the $F-$function gives a correct
result for the isolated finite systems even for small number of interacting
particles (provided the number of principal components in the eigenstates is
large).

The advantage of our approach, in comparison with direct calculations of
complex quantum systems, is that we do not need to diagonalize huge
Hamiltonian matrices in order to perform calculations of observables for
excited states. Indeed, for the full statistical description of such
systems, one needs to know the average shape of the compound states (rather
than the eigenstates themselves) and the unperturbed energy spectrum.
Therefore, the problem of analytical description of the shape of eigenstates
is the central point in the technical implementation of the approach. One
should note that the average shape of eigenstates is the same as that of the
local spectral density of states (LDOS), if the interaction is not extremely
strong.

For small interaction the shape of the chaotic eigenstates is known to be
well described by the Breit-Wigner form. However, in practice, this region
is small if the number of particles is not very large. With an increase of
the interaction strength, the average shape of the eigenstates ($F-$%
function) changes from the Breit-Wigner one to that close to the Gaussian
with the exponential tails. As we have found (Appendix 1), the correct
description of the shape requires two essential parameters. The first one
is, in fact, the half-width of the $F-$function which is close to the
Breit-Wigner half-width and for weak interaction is given by the Fermi
Golden Rule. Another parameter is defined by the root-mean-square width of
the $F-$function, or, the same, by an effective band-width of the
Hamiltonian matrix in the energy representation.

In this paper we suggest the phenomenological expression for the $F-$%
function which is valid in a large region of the interaction strength and
other parameters. This expression allows for the analytical and numerical
calculations of different mean values and transition amplitudes. As one
example, we have derived analytical expression for the distribution of the
occupation numbers $n_s(E)$ in isolated systems of $n$ interacting
Fermi-particles distributed over $m$ orbitals (we have used the simple
Gaussian approximation of the $F-$function, see Section 5).

By making use of the $F-$function, we have also studied the validity of the
standard Fermi-Dirac distribution for the description of finite systems of
interacting Fermi-particles. As was found, the Fermi-Dirac distribution can
provide reasonable approximation for both isolated and open (in the thermal
bath) systems. However, the parameters of the Fermi-Dirac distribution have
to be re-defined by taking into account the increase of the effective
temperature which is due to the effects of the interaction. We calculated
this increase of the temperature analytically and compared with the
numerical experiments for the two-body random interaction model; the data
show very good agreement.

One should stress that our approach gives more accurate result for $n_s$ and
has much wider region of the applicability, compared to the Fermi-Dirac
distribution. Specifically, it is valid even in the region where the
Fermi-Dirac distribution fails, for example, due to the small number of
particles.

To conclude with, we would like to point out that similar approach may be
also used for classical chaotic systems. Indeed, let us consider the system
described by the Hamiltonian $H=H_0+V$ where $H_0$ is the unperturbed
(``simple'') Hamiltonian and $V$ stands for ``complex'' interaction between
particles. Assume that we know the distribution for some variable in the
system described by $H_0$ , for example, the statistical average for the
energy distribution of a single particle $n(\epsilon ,E_0)$ where $\epsilon $
is the energy of the particle and $E_0$ is the total energy of the system.
Then, one can calculate the effect of the interaction $V$ by averaging of $%
n(\epsilon ,E_0)$ over the unperturbed energy $E_0$ using the $F-$function, $%
n(\epsilon ,E)=\int n(\epsilon ,E_0)F(E,E_0)dE_0$. Here $F(E,E_0)$ gives the
probability of different values of $E_0$ for a given value of $E$. As was
indicated in \cite{CCGI96} and checked numerically in \cite{BGIC97,WIC97},
this classical $F-$distribution coincides with the shape of quantum
eigenstates in the semiclassical region which turns out to be very wide.
Thus, the knowledge of the shape of quantum eigenstates can be used for the
classical calculations and vice versa.

{\bf Acknowledgements}

The authors are thankful to Y.Fyodorov, G.Gribakin, M.Kuchiev, J.-L.Pichard,
I.Ponomarev, V.Sokolov and O.Sushkov for the discussions; F.M.I is very
grateful to the staff of School of Physics, University of New South Wales
for the hospitality during his visit when this work was done. This work was
supported by the Australian Research Council. F.M.I. acknowledges the
partial support by Grant No.94-2058 from INTAS.

{\bf Appendix 1. Structure of chaotic eigenstates and spreading function}

For practical implementation of the above approach one needs to know the
average shape of compound eigenstates (the $F-$function). One should stress
that there is no simple analytical expression valid in a large range on the
interaction strength $V$ . For example, the popular Breit-Wigner expression
(Lorentzian) is not good for obvious reason: it has infinite second moment.
The question of an appropriate description of chaotic eigenstates in
realistic many-body systems has been studied in detail for Ce atom \cite
{FGGK94}. In particular, it was found that good correspondence to the
numerical data is given by two phenomenological expressions. The first one
is $F(x)\sim \exp (-\sqrt{1+4x^2})$ where $x\,=(E_k-E)/\Gamma $ with $E_k$
as the energy of a basis state $\left| k\right\rangle $ and $\Gamma $ as the
effective width of the distribution. This expression is close to the
Gaussian at the central part and is exponential in the tails (see similar
conclusions in \cite{zel2} where nuclear shell model was studied).

Another expression which is more convenient for the analytical study is the
so-called ``squared'' Lorentzian \cite{FGGK94},

\begin{equation}
\label{bw2}F(E_k-E)\sim \frac 1{\left[ \left( E_k-E\right) ^2+\frac{\Gamma ^2%
}4\right] ^2};\,\,\,\,\,\,E\,=E^{(i)}+\Delta _1^{(i)} 
\end{equation}
Here $\Delta _1^{(i)}\ll \Gamma $ is some small shift (see below) which in
the zero approximation can be neglected, and $E_k$ is defined by 
\begin{equation}
\label{hkk}E_k=H_{kk}=\sum\limits_sn_s^{(k)}\,\epsilon
_s+\sum\limits_{s>p}u_{sp}n_s^{(k)}n_p^{(k)} 
\end{equation}

Since the ``resonant'' dependence $E_k-E^{(i)}\,$of the spreading function $%
F_k^{(i)}$ for not extremely strong interaction is symmetric in indexes $i$
and $k$ , the value of $\Gamma =2\sqrt{\overline{\left( \Delta E\right) ^2}}$
can be expressed in terms of the second moment of $F\,$ using the following
exact relation for the basis components, see Appendix 2, 
\begin{equation}
\label{delE}(\Delta E)^2\equiv \sum\limits_i\left| C_k^{(i)}\right| ^2\left(
E_k-E^{(i)}\right) ^2=\sum\limits_{p\neq k}H_{kp}^2 
\end{equation}
with $H_{kp}$ standing for non-diagonal Hamiltonian matrix elements defined
by the residual interaction $V$. This allows us to find the second moment of
the spreading function $F(E)\,.\,$For example, in the case of $n$ particles
distributed over $m$ orbitals we have \cite{FGI96} 
\begin{equation}
\label{50}\left( \frac \Gamma 2\right) ^2=\overline{\left( \Delta E\right) ^2%
}=\frac{V^2}4n(n-1)(m-n)(3+m-n) 
\end{equation}
Here $V^2=\,\overline{\left| V_{st\rightarrow pq}\right| ^2}$ is the mean
squared value of non-diagonal matrix elements of the two-body residual
interaction.

Our detailed study of the two-body random interaction model \cite
{FGI96,FIC96,FI97} has revealed that the shape of the eigenfunctions. (as
well as the local spectral density of states) strongly depends on the
relative strength of the interaction. Namely, with an increase of the
interaction $V$ , the shape of the $F-$function changes its form from the
Breit-Wigner one to the nearly Gaussian (see also \cite{CFI97}). It was
found that for small residual interaction the shape of eigenstates has more
complicated form,compared to (\ref{bw2}), and should be characterized by two
different widths. Indeed, the half-width of the $F-$distribution is given by
the Fermi golden rule $\Gamma _{BW}=2\pi \overline{v^2}/d_f$ where $v\,$is
the matrix element of the residual interaction coupling a particular basis
component with other basis states $\left| f\right\rangle $ directly coupled
by the two-body interaction, and $d_f$ is the energy spacing between these
basis states (see details in Appendix 3). On the other hand, there is the
relation (\ref{50}) which defines another width $\Gamma =2\sqrt{\overline{%
(\Delta E)^2}}$ via the second moment. One should stress that these two
widths are parametrically different in the interaction, $\Gamma _{BW}\sim
V^2 $ , and $\Gamma \sim V$ . There is also the ``non-resonant'' energy
dependence $F\propto \rho ^{-1}$ (slow variation of the $F-$function due to
the change of the density of states $\rho (E))\,$ which should be taken into
account. This dependence follows from the estimate $F_{max} \sim N_{pc}^{-1}
\sim \Gamma \rho $.

The above arguments allow us to find more universal expression for the
spreading function $F$ when $\Gamma _{BW}<\Gamma $ \cite{FI97},

\begin{equation}
\label{newBW}F_k^{(i)}\sim \frac{\left( \rho _0(E_k)\rho (E^{(i)})\right)
^{-1/2}}{\left[ \left( E_k-E\right) ^2+\frac{\Gamma _1^2}4\right] \left[
\left( E_k-E\right) ^2+\frac{\Gamma _2^2}4\right] } 
\end{equation}
Here we take into account the shift of the maximum of the $F-$function by
the relation $E=E^{(i)}+\Delta _1^{(i)}$ . The two parameters, $\Gamma _1$
and $\Gamma _2$ are directly related to the above two widths, $\Gamma
_1=\Gamma _{BW}\,\,$and$\,\,\Gamma _2=\Gamma ^2/\Gamma _1$. The value of $%
\Gamma _2$ is found from the relation $\sum%
\nolimits_iF_k^{(i)}(E^{(i)}-E_k)^2=\ \overline{(\Delta E)^2}=\Gamma ^2/4$
(see (\ref{50}) and Appendix 2) by integration of (\ref{newBW}) in the
approximation $\rho =const$. Here and below we assume that the $F-$ function
is normalized, $Z=1$. In the expression (\ref{50}), $\rho _0(E_k)$ is the
density of basis (unperturbed) states and $\rho (E^{(i)})\,$ is the density
of compound states. We assume they are smooth functions of the energy and
this energy dependence is slow in comparison with the ``resonant'' energy
dependence on the scale $\Gamma $. Symmetric dependence on $\rho _0$ and $%
\rho $ has been chosen in order to keep symmetry in indexes $k$ and $i$ in
the $F-$function.

For very small $V\,$ we have $\Gamma _1\ll \Gamma _2$ (also, $\Delta
_1^{(i)}\ll \Gamma _1$, see below) , therefore, in the central part the $F-$%
distribution (\ref{newBW}) has the Breit-Wigner shape with the width $\Gamma
_{BW}$ . Concerning the meaning of $\Gamma _2$ , it is the effective energy
band width of the Hamiltonian matrix $H$ , which is due to the two-body
nature of the interaction. Indeed, the expression for $\Gamma _2$ is given
by the estimate 
\begin{equation}
\label{Gamma2}\Gamma _2=\frac{\Gamma ^2}{\Gamma _{BW}}\approx \frac{d_f}{%
2\pi }n(n-1)(m-n)(m-n+3)\approx d_0(m-n) 
\end{equation}
which is independent of the interaction strength $V$ . Here we have used
expression (\ref{50}) and the estimate of the average value $d_f\equiv
d_0/M_f$ for high excited states ($M_f$ here is the normalized density of
those basis states which are directly connected to the chosen state, see
details in Appendix 3). On the other hand, the typical band width $\Delta
_H\equiv 2bd_0\,$of the two-body interaction Hamiltonian matrix is about
four times, $\Delta _H\approx 4(m-n)d_0$ , of the energy needed to transfer
the particle from the Fermi-level $\epsilon _F=nd_0$ to the highest
available orbital $\epsilon _m=md_0\,.$ Therefore, the estimate for $\Gamma
_2$ reads as $\Gamma _2\approx \Delta _H/4 $ .

Now, we can easily explain the form of the $F-$function (\ref{newBW}) using
the perturbation theory in the interaction $V$ . First, let us consider the
energy interval $\Gamma _{BW}<\left| E_k-E^{(i)}\right| <\Delta _H$ . Within
this interval, the basis state $\left| k\right\rangle $ can be coupled to
the principal components of the state$\left| i\right\rangle $ in the first
order of $V$ , $\left| C_k^{(i)}\right| ^2\sim \left( \frac{V_{ik}}{%
E_k-E^{(i)}}\right) ^2$ . This quadratic decay agrees with the Breit-Wigner
shape of the $F-$function. Outside the energy band $\Delta _H$ , for $\Delta
_H\,$$<\left| E_k-E^{(i)}\right| <2\Delta _H$ , the basis state $\left|
k\right\rangle $ can be coupled to the principal components of the state$%
\left| i\right\rangle $ in the second order of $V$ $,$ resulting in the
dependence $\left| C_k^{(i)}\right| ^2\sim \left( \frac{V_{ik}}{E_k-E^{(i)}}%
\right) ^4$ . This corresponds to the tails of the squared Lorentzian shape (%
\ref{newBW}). Therefore, our expression (\ref{newBW}) seems to be good in a
large energy interval and the second moment is finite which is important for
applications. And finally, longer tails are described by higher orders $\nu =
\frac{\left| E_k-E^{(i)}\right| }{\Delta _H}$ of the perturbation theory,%
$$
\left| C_k^{(i)}\right| ^2\sim \left( \frac{V_{ik}}{E_k-E^{(i)}}\right)
^{2\nu } 
$$
\begin{equation}
\label{high}=\exp \left( -\frac{2\left| E_k-E^{(i)}\right| }{\Delta _H}\ln
\left( \frac{\left| E_k-E^{(i)}\right| }V\right) \right) 
\end{equation}
This explains the exponential tails of the $F-$function, see details in \cite
{FGGK94}.

Numerical calculations \cite{FGGK94,ZELE,zel2,CFI97} demonstrate that for
stronger interaction $V$ , the width of the spreading function $F$ rapidly
becomes linear in $V$ (instead of the quadratic dependence in $\Gamma _{BW}$
) and it is better to use (\ref{bw2}). One can write the extrapolation
expression both for small and large values of $V$ (see also \cite{zel2}): 
\begin{equation}
\label{inter}\Gamma _1=\frac{\Gamma _{BW}\Gamma }{\Gamma _{BW}+\Gamma } 
\end{equation}
As a result, for small $V$ we have $\Gamma _{BW}\ll \Gamma $ and $\Gamma
_1=\Gamma _{BW}\sim V^2$ and for larger values $\Gamma _1\approx \Gamma
_2\approx \Gamma \sim V\gg \Delta _1^{(i)}$ . The critical value for this
transition is given by the relation $\Gamma _{BW}=\Gamma =\Gamma _2\sim
\Delta _H$ and reads as $V_{cr}\approx d_f\,n(m-n)/(2\pi )$ , see also (\ref
{50}). The estimate of the average value of $d_f\equiv d_0/M_f$ (see
Appendix 3) far from the ground state gives $V_{cr}\sim d_0/n$ . As was
discussed in Section 9, the equilibrium distribution occurs for $V>d_f$ ,
this results in a quite unexpected conclusion. Namely, the validity of the
standard Breit-Wigner shape turns out to be very strongly limited since the
region $1<V/d_f\ll V_{cr}/d_f\approx \,n(m-n)/(2\pi )$ is practically absent
for small number of particles $n$ and orbitals $m$ .

The shift $\Delta _1^{(i)}$ in (\ref{newBW}) stands due to the level
repulsion which forces eigenvalues $E^{(i)}$ in the lower half of the
spectrum to move down. The mean field energies $E_k=H_{kk}$ do not include
the non-diagonal interaction which leads to the repulsion, therefore, the
``center'' of the $F-$function is shifted by the value $\Delta
_1^{(i)}=H_{ii}-E^{(i)}$ where $H_{ii}$ is the diagonal matrix element of
the Hamiltonian matrix. This shift $\Delta _1^{(i)}$ can be estimated from
general arguments. Indeed, the shape of the density of states is the same
for both interacting and non-interacting particles \cite{FW70,BF71}, with
the same position $E_c\,$of the centers of $\rho _0(E)$ and $\rho (E)$ (due
to the conservation of the trace of the Hamiltonian $H$ ). However, the
variances of $\rho _0$ and $\rho $ are different. This means that one can
use the scaling relation $(D\rightarrow KD)$ for the energy intervals $D$
and find the scaling coefficient $K$ from the relation between the
variances, $\sigma ^2=\sigma _0{}^2+\overline{(\Delta E)^2}=K^2\sigma _0^2$
where $\overline{(\Delta E)^2}$ is defined by (\ref{50}). Since the center $%
E_c\,$ for the energy level density $\rho (E)$ does not shift, one can
obtain the following shift of the levels: 
\begin{equation}
\label{center}\Delta _1^{(i)}=\left( E_c-E^{(i)}\right) \left( \sqrt{1+\frac{
\overline{\ (\Delta E)^2}}{\sigma _0^2}}-1\right) 
\end{equation}
The value of $\overline{(\Delta E)^2}$ is typically much less than $\sigma
_0{}^2$ , therefore, one can get

\begin{equation}
\label{shift1}\Delta _1^{(i)}\simeq \left( E_c-E^{(i)}\right) \frac{
\overline{\ (\Delta E)^2}}{2(\sigma _0)^2} 
\end{equation}

Another way to obtain the shift $\Delta _1^{(i)}$ is related to the exact
relation for the first moment of $F$ , see Appendix 2, 
\begin{equation}
\label{ek}E_k=\sum\limits_iE^{(i)}F_k^{(i)}\approx \int F_k^{(i)}\rho
(E^{(i)})E^{(i)}dE^{(i)} 
\end{equation}
The substitution of the expression (\ref{newBW}) into (\ref{ek}) results in
the following value of the shift \cite{FI97},

\begin{equation}
\label{shift}\Delta _1^{(i)}\simeq \frac 12\frac{d(ln\,\rho )}{dE}\overline{%
\left( \Delta E\right) ^2} 
\end{equation}
According to \cite{FW70,BF71} (see also \cite{brody}) the shape of density
of states for $m\gg n\gg 1$ is close to the Gaussian both for
non-interacting and interacting particles with $E_c$ and $\sigma _0^2$ as
the center and the variance of the energy distribution $\rho _0(E)$
(respectively, $\sigma ^2$ for $\rho (E)$). In this case the relation (\ref
{shift}) gives the same estimate (\ref{shift1}) for the shift $\Delta
_1^{(i)}$.

The fact that the two different derivations of the shift $\Delta _1^{(i)}$
lead to the same result is far from being trivial since the assumptions for
the two derivations of $\Delta _1^{(i)}$ are different. Indeed, the second
derivation of (\ref{shift1}) is based on the specific dependence of the
eigenstate shape on the densities $\rho _0$ and $\rho ,$ unlike the general
derivation of (\ref{center}). One should stress that the specific form of
the ``resonant'' energy dependence of the $F-$function (the denominator in (%
\ref{newBW}) defined by the squared Lorentzian, Gaussian, etc.) is not
important for Eq. (\ref{shift}) provided $\overline{\left( \Delta E\right) ^2%
}$ is fixed. The only assumption in the above derivation is the possibility
to expand the density $\rho (E)$ near the maximum of the $F-$function. In
fact, above we have demonstrated that the non-resonant ``distortion'' factor 
$\xi \equiv $$\left( \rho _0(E_k)\rho (E^{(i)})\right) ^{-1/2}$ in (\ref
{newBW}) is necessary.

Thus, the phenomenological expression for the shape of the $F-$function (\ref
{newBW}) is self-consistent. Note, that one can use other expressions for
the $F-$function (see, for example, \cite{CFI97}), however, it should
contain both the resonance term depending on $E_k-E^{(i)}$ and the density
distortion factor $\xi $ .

{\bf Appendix 2. Moments of the $F-$function and energy spectrum}

Here we calculate the first and the second moment of the function $F_k^{(i)}$
over the perturbed spectrum $E^{(i)}$ . Note, that the dependence of $%
F_k^{(i)}$ on the energy $E^{(i)}$ is known as the local spectral density of
states (LDOS), or the ``strength function''. On definition,%
$$
\left\langle E^{(i)}\right\rangle _k=\sum_iE^{(i)}F_k^{(i)}\approx
\sum_i\left| \left\langle k,i\right\rangle \right| ^2E^{(i)} 
$$
\begin{equation}
\label{Ekk}=\sum_{i,j}\left\langle k|i\right\rangle \left\langle i\right|
H\left| j\right\rangle \left\langle j|k\right\rangle =H_{kk}=E_k
\end{equation}
where the relation $\left\langle i\right| H\left| j\right\rangle =\delta
_{ij}\,\left\langle i\right| H\left| i\right\rangle $ is used for the exact
eigenstates. The variance can be obtained using the matrix elements of $H^2$%
$$
\overline{\left( \Delta E\right) _k^2}=\sum\limits_iF_k^{(i)}\left(
E_k-E^{(i)}\right) ^2\approx \sum_i\left| C_k^{(i)}\right| ^2\left(
E_k-E^{(i)}\right) ^2 
$$
\begin{equation}
\label{deelEN}=\sum\limits_{p\neq k}H_{kp}^2
\end{equation}
For example, in the two-body random interaction model with $n$ particles
distributed over $m$ orbitals the sum in (\ref{deelEN}) can be evaluated
exactly (see (\ref{50})).

Now, we calculate the first moment and variance of the energy spectrum. The
trace conservation of $H$ gives the first moment,%
$$
E_c=\frac 1N\sum_kH_{kk}-\left( \frac 1N\sum_kH_{kk}\right) ^2 
$$
The conservation of $TrH^2$ gives%
$$
\overline{E^2}=\frac 1N\sum_k\left\langle k\right| H^2\left| k\right\rangle
=\frac 1N\sum_{k,p}\left| H_{kp}\right| ^2 
$$
which results in the relation%
$$
\sigma ^2\equiv \overline{E^2}-E_c^2=\frac 1N\sum_{p,k}H_{pk}H_{kp}-E_c^2 
$$
$$
=\frac 1N\sum_kH_{kk}^2+\frac 1N\sum_{p\neq k}H_{pk}^2-E_c^2 
$$
$$
=\frac 1N\sum_kE_k^2-E_c^2+\frac 1N\sum_k\overline{\left( \Delta E\right)
_k^2}=\sigma _0^2+\overline{\left( \Delta E\right) ^2} 
$$
where $\sigma _0^2$ is the variance of the unperturbed spectrum and we have
used Eq.(\ref{deelEN}).

{\bf Appendix 3. Calculations of spreading widths}

To start with, we should stress that there are different definitions of the
spreading widths. One of the natural definitions is $\Gamma _k\equiv 2\sqrt{
\overline{\left( \Delta E\right) _k^2}}$ where $\left( \Delta E\right) _k^2$
is the variance of the distribution of the components $F_k^{(i)}$ (see (\ref
{deelEN}))$.$ In Ref.\cite{FGI96} it was shown that in the model of random
two-body interaction this quantity is constant, $\Gamma _k=\Gamma $ , see (%
\ref{50}), i.e., it does not depend on a particular basis state (therefore,
on excitation energy, number of excited particles $n^{*}$ corresponding to
this state, etc.). Note, it has a linear dependence on the interaction
strength $V$.

Unlike the latter, the commonly used definition of the spreading width is
related to the Breit-Wigner distribution and is defined as its half-width $%
\Gamma _{BW}$ . However, this definition is reasonable only for relatively
small interaction, when the form of the $F-$ function is indeed close to the
Breit-Wigner form. In this case, the spreading width is given by the Fermi
Golden Rule, 
\begin{equation}
\label{GamBW}\Gamma _{BW}^{(k)}=2\pi \frac{\overline{V_{kf}^2}}{d_f} 
\end{equation}
where $d_f\,$ is the mean spacing between corresponding basis states $\left|
f\right\rangle $ and $V_{kf}\,\,$is the matrix element of the interaction
between the basis states $\left| k\right\rangle $ and $\left| f\right\rangle
\,\cite{comment}.$ As one can see, the spreading width $\Gamma _{BW}^{(k)}$
is proportional to $V^2$ and differs from $\Gamma $ . Note, that the second
moment of the Breit-Wigner shape diverges, however, actual form has always
cut-off in the tails which is reflected by the finite value of $\Gamma $ .

If the interaction $V$ is not small, the form of the $F-$function
significantly differs from the Breit-Wigner shape (see Appendix 1). The
critical value of $V$ for this transition can be estimated from the
condition $\Gamma _{BW}^{(k)}$ $\approx $ $\Gamma _k$ .

Contrary to the spreading width $\Gamma \,,$ the half-width of the
Breit-Wigner shape depends on the basis state $\left| k\right\rangle \,.$
Let us start with the estimate of the mean value of the spreading width $%
\Gamma _{BW}^{(k)}$ . According to the definition (\ref{GamBW}), one needs
to calculate the density of squared transition matrix elements $\overline{%
V_{kf}^2}/d_f=\left( \sum_fV_{kf}^2\right) /\Delta $ where the sum is taken
over transitions from a given basis state $\left| k\right\rangle $ to other
basis states $\left| f\right\rangle $ in the energy interval $\Delta $ . One
should stress that the number of ``allowed'' transitions (due to the
two-body interaction) is much less than the total number of basis states in
this interval $\Delta $ . Let us assume, for simplicity, that the spacings $%
d_0$ between single-particle energy levels is constant and consider a pair
of particles which occupy the orbitals $s$ and $q$ . If we move one particle
to higher orbital and another particle to lower orbital by the same energy
interval, the total energy of particles does not change. The total number of
such moves is $M\approx (\min (m-s,q)+\min (m-q,s))/2$ where $m\gg 1$ is the
number of orbitals. By averaging over all values $s$ and $q$ , one can
obtain $M\approx m/3$ . The Pauli principle reduces the number of available
orbitals, therefore, in the system with $n$ particles we have $M\approx
(m-n)/3$ . The number of possible pairs is given by $n(n-1)/2$ , thus, the
total number of basis states $\left| f\right\rangle $ which have the same
energy and connected with the chosen basis state $\left| k\right\rangle $ is
defined by 
\begin{equation}
\label{Mff}M_f\approx (m-n)n(n-1)/6\, 
\end{equation}
Other basis states are separated by, at least, the energy distance $d_0$ .
As a result, we obtain 
\begin{equation}
\label{GammaAV}\left\langle \Gamma _{BW}\right\rangle \approx 2\pi M_f\frac{%
V^2}{d_0}\approx \frac \pi 3(m-n)n(n-1)\frac{V^2}{d_0} 
\end{equation}

It is interesting to note that the simple estimate involving total strength
of transitions $\overline{\left( \Delta E\right) ^2}$, see Eq.(\ref{50}),
divided by the band-width $\Delta \sim d_0(m-n)$ , gives close result. The
above estimate (\ref{GammaAV}) can be used when studying the shape of the $%
F- $ function in the regime of weak interaction. Note, the value of $M_f$
defines the mean spacing between the ``allowed'' final states is 
\begin{equation}
\label{dff}d_f=d_0/M_f 
\end{equation}

One should stress that the above estimate of $M_f$ has been obtained for the
spreading width averaged over all basis states. Near the ground state actual
value of $M_f$ is much smaller due to limitation of the available phase
space. Also, the spreading width depends on the number of excited particles
in the basis state. For example, one can calculate the spreading width $%
\Gamma _{BW}^1$ of the basis states with one excited particle only, 
\begin{equation}
\label{GamBW1}M_f=(s-n)^2;\,\,\,\,\Gamma _{BW}^1\approx \pi \frac{%
(s-n)^2V_0^2}{d_0};\,\,\,\,\,\,\,\,\,s\leq 2n
\end{equation}
\begin{equation}
\label{GamBW2}M_f=n(2s-3n);\,\,\,\Gamma _{BW}^1\approx \pi \frac{%
n(2s-3n)V_0^2}{d_0};\,\,\,s>2n
\end{equation}
Here $s\,$ is the position of a particle corresponding to the energy $%
\epsilon _s\approx s\,d_0$ .

FIGURE CAPTIONS\\

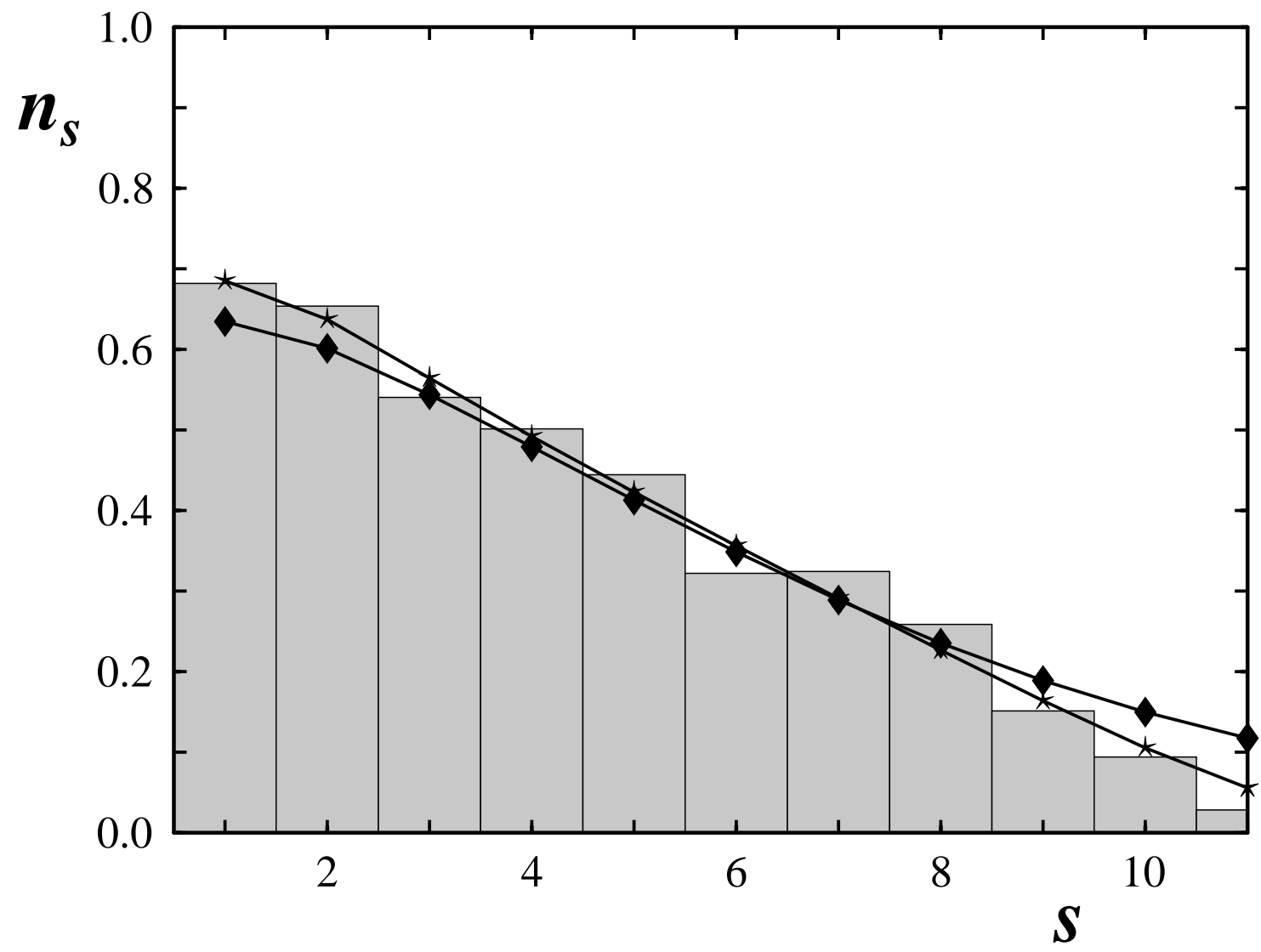
Fig.1. Analytical description of the occupation numbers. The data are given
for the two-body random interaction model (\ref{H}) of $n=4$ Fermi-particles
distributed over $m=11$ orbitals with $V=0.20$ and $d_0=1$ in the definition
of single-particle energies, $\epsilon _s=d_0(s+\frac 1s),$ see \cite
{FIC96,FGI96,FI97}. The histogram is obtained according to (\ref{ns}) by the
averaging over eigenstates with energies taken from small energy window
centered at $E=17.33$ and over $20$ Hamiltonian matrices (\ref{H}) with
different realization of the random interaction. Stars represent the
analytical expression (\ref{R}) with $\sigma _{0s}$ found from
single-particle energy spectrum. Diamonds correspond to the Fermi-Dirac
distribution with thermodynamical temperature (\ref{Ttherm}) and chemical
potential found from the standard condition for the total number of
particles, $n=\sum_s n_s$ .

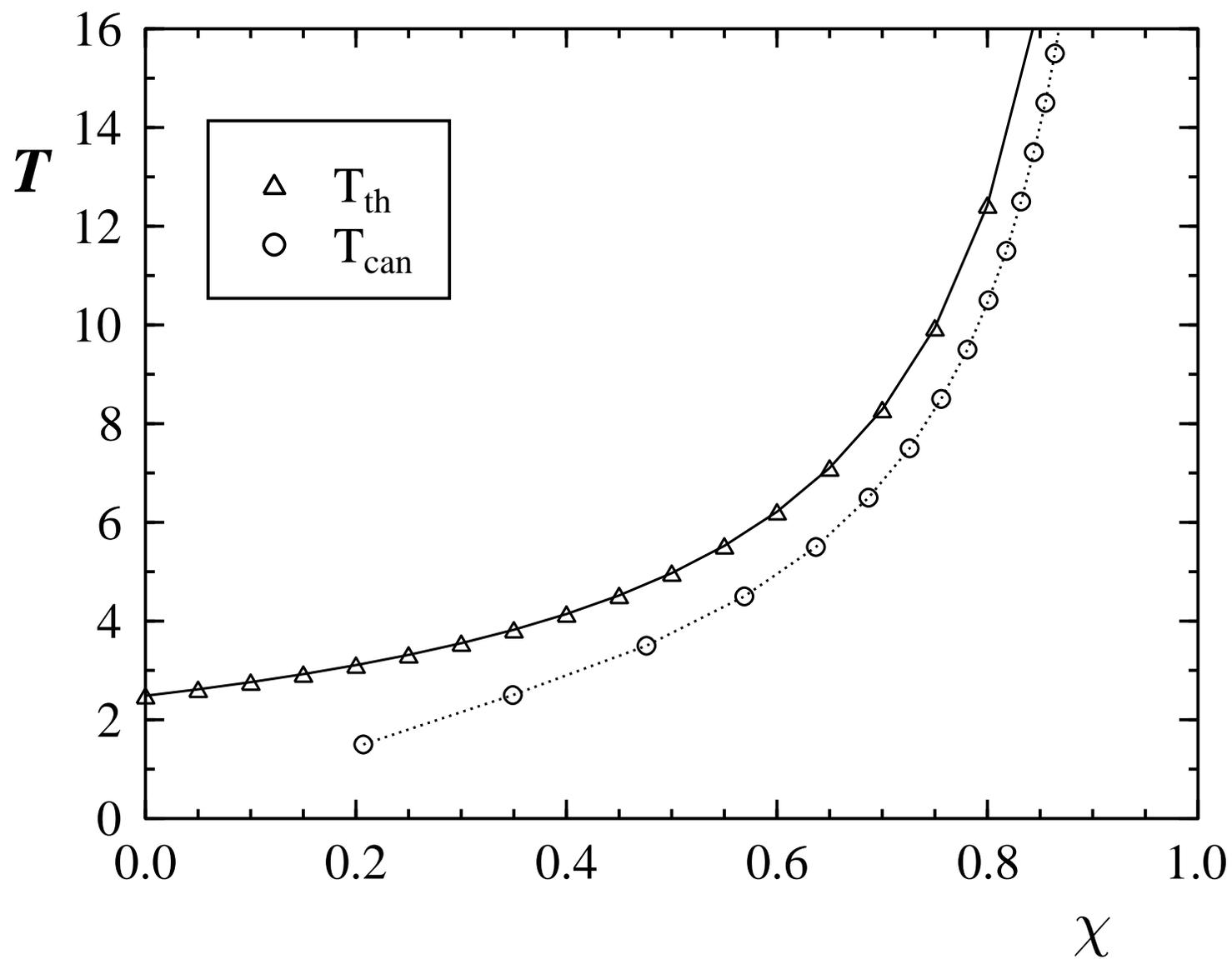
Fig.2. Different temperatures versus the rescaled energy $\chi
=(E-E_{fermi})/(E_c-E_{fermi})$ for the two-body random interaction model
with $n=4$ Fermi- particles and $m=11$ orbitals. Triangles stand for the
thermodynamical temperature $T_{th}$ defined by (\ref{Ttherm}) and should be
compared to the canonical temperature $T_{can}$ (circles), see (\ref{Tcan}).
The width $\sigma $ of the perturbed density of states is defined by the
residual interaction $V=0.12$ according to (\ref{sigma}) and (\ref{50}) with 
$\sigma _0$ found numerically from the unperturbed many-particle energy
spectrum (the mean level spacing $d_0$ between single-particle levels is set
to $d_0=1$).

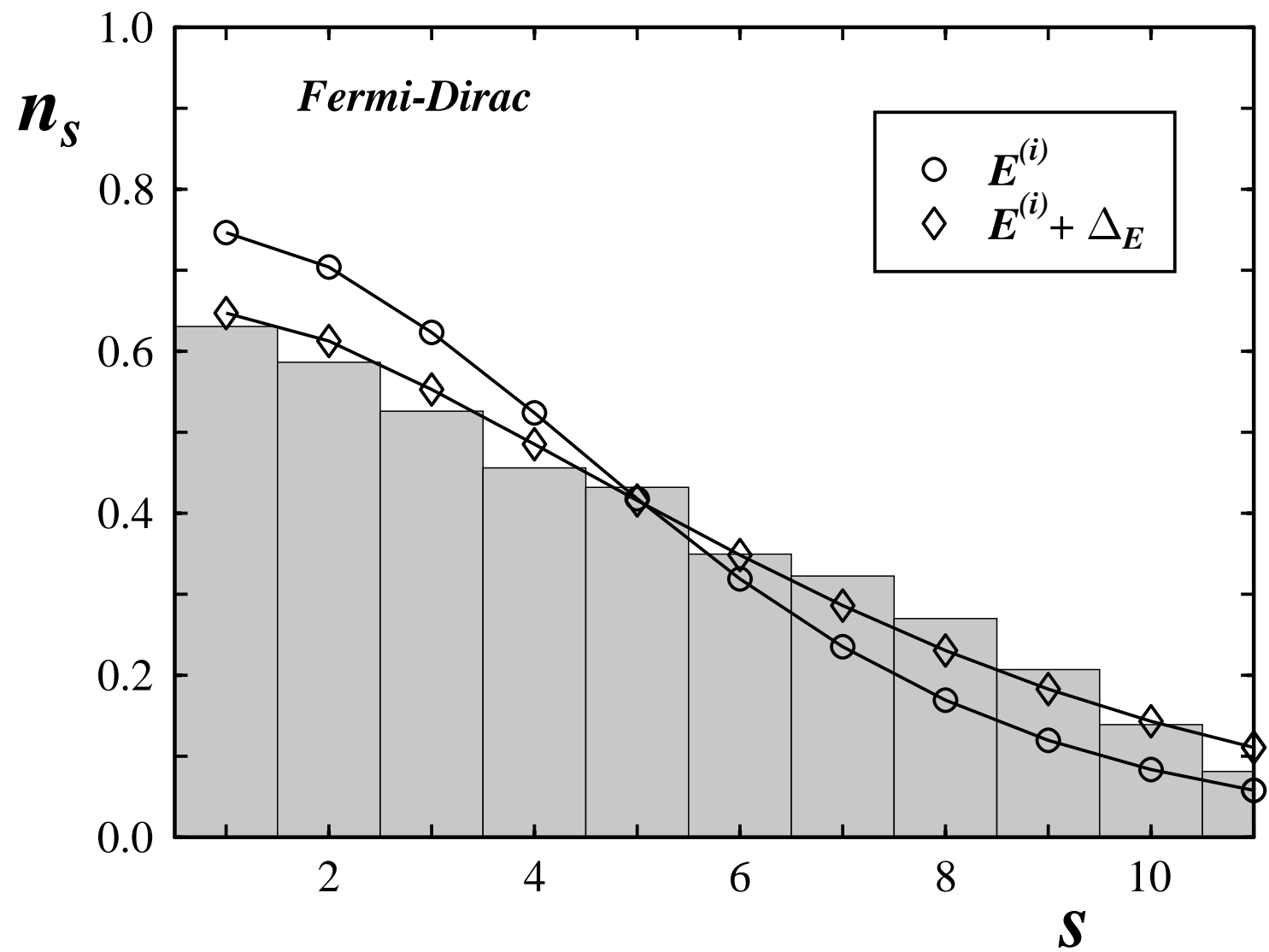
Fig.3. Fermi-Dirac distribution for strongly interacting particles. The data
are given for the two-body random interaction model (\ref{H}) with the
parameters of Fig.1 (the rescaled energy is $\chi=0.55$). Circles stand for
the Fermi-Dirac distribution with the total energy $E$ corresponding to the
energy of eigenstates, see (\ref{eqs}). Diamonds correspond to the shifted
energy according to the expression (\ref{Delfinal}).

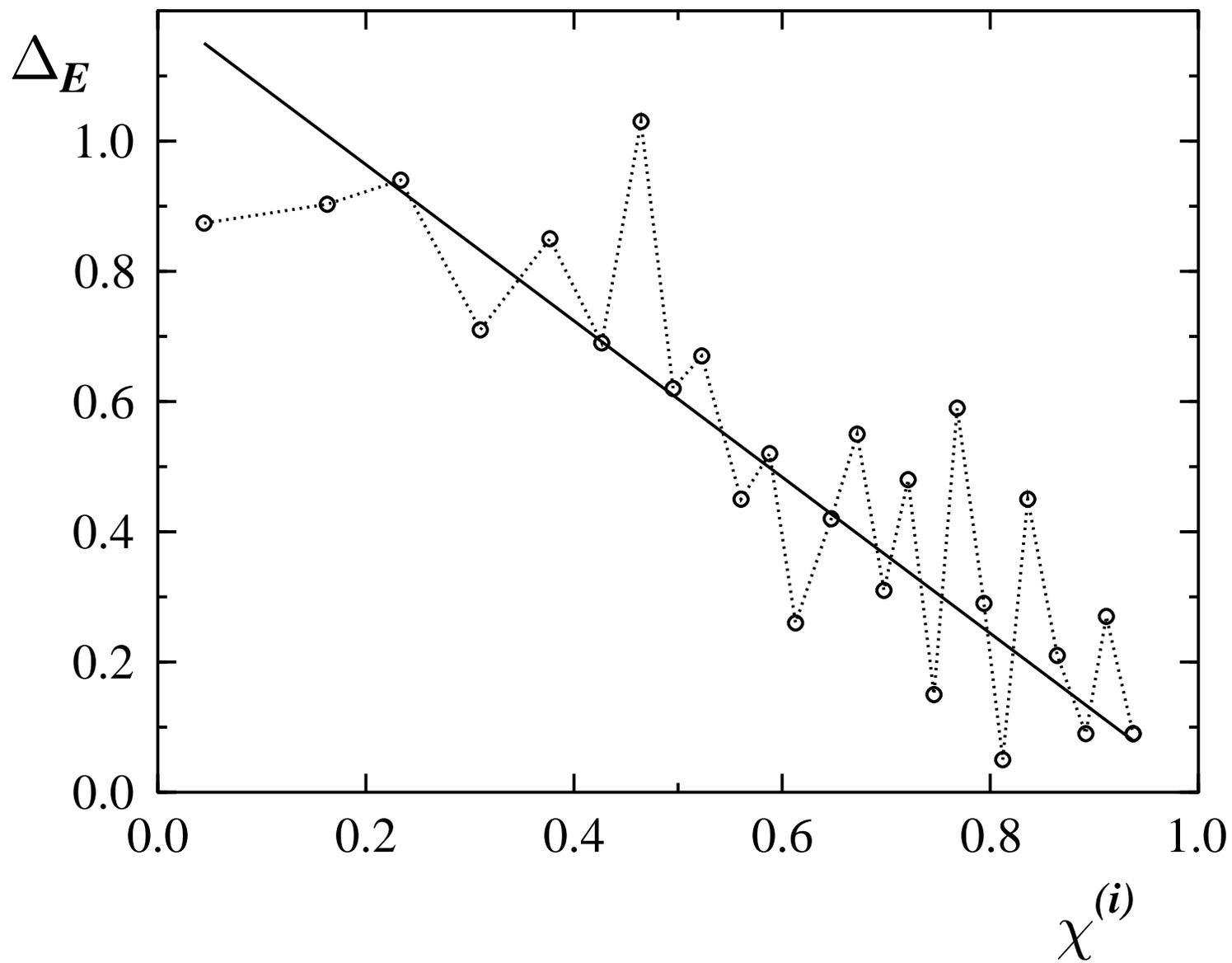
Fig.4. Shift of the total energy for the corrected Fermi-Dirac distribution.
The data are given for the model (\ref{H}) with $n=4,m=11,d_0=1,V=0.12$, see
explanation in the text. The straight line is the analytical expression (\ref
{Delfinal}); the dotted line (circles) present direct computation of the
shift based on the diagonalization of the Hamiltonian (\ref{H}) with the
following computation of the $<E_k>_i$ . On the horizontal axes the rescaled
energy $\chi ^{(i)}=(E^{(i)}-E_{fermi})/(E_c-E_{fermi})$ is plotted.

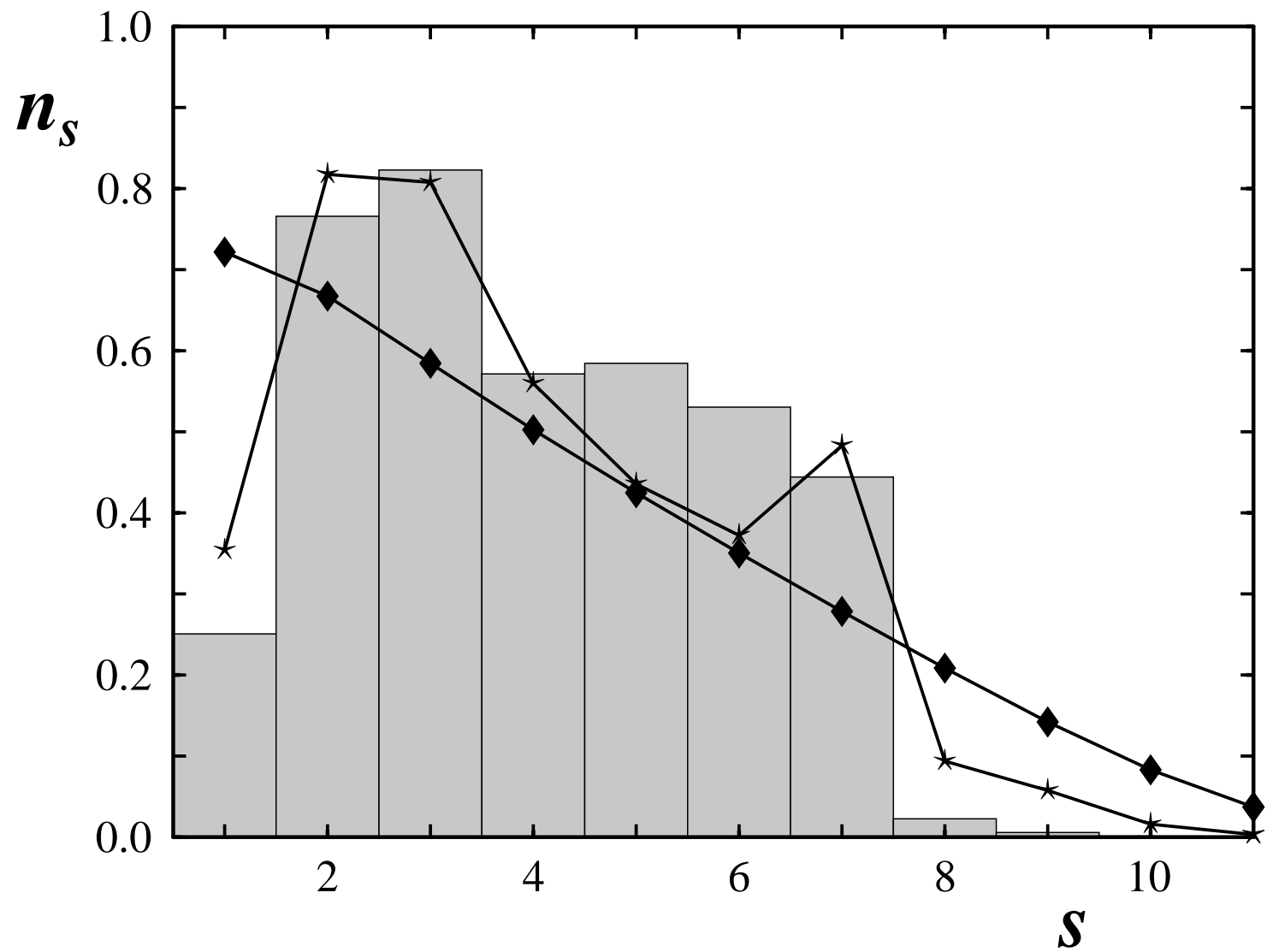
Fig.5. Distribution of the occupation numbers for small interaction. The
histogram is obtained in the same way as in Figs.1-2, for the very weak
interaction $V=0.02$ which correspond to the region of (II) of the ``initial
chaotization'', see Section 10. The total energy (center of the small energy
window) is $E^{(i)}=17.33$. Diamonds correspond to the theoretical
expression (\ref{R}) which is not valid in this region due to absence of
equilibrium. Stars are obtained by direct numerical computation of $n_s$
according to the F-distribution (\ref{nalpha}) with the $F-$function taken
in the form (\ref{newBW}), see Appendix 1 and \cite{FI97}. The latter values
are closer to the ``experimental'' ones since we performed summation over
real unperturbed spectrum (instead of the integration with the Gaussian
approximation for $\rho _o$ used to derive Eq. (\ref{R})).


\begin{thebibliography}{99}
\bibitem{comment}  We would like to point out that an alternative definition
involves exact eigenstates $\left| i\right\rangle $ (instead of basis state $%
\left| f\right\rangle )$ of the Hamiltonian matrix with the excluded basis
state $\left| k\right\rangle $ (see, for example, \cite{BM69,ZELE}); for
small interaction this definition gives the same result.
\end{thebibliography}
\end{document}